\documentclass[a4paper,12pt,twoside]{article} 
\usepackage{a4wide}
\usepackage{graphicx}
\usepackage{rotating}
\usepackage{cite}
\usepackage[reqno]{amsmath}
\usepackage{epsfig}
\usepackage{array}
\usepackage{float}
 
\def\ra{\rightarrow} 
\newcommand{\beq}{\begin{equation}}
\newcommand{\eeq}{\end{equation}}
\def\gs{\mathrel{ \rlap{\raise
0.511ex \hbox{$>$}}{\lower 0.511ex \hbox{$\sim$}}}} \def\ls{\mathrel{
\rlap{\raise 0.511ex \hbox{$<$}}{\lower 0.511ex \hbox{$\sim$}}}}

\newcommand{\ba}{\begin{array}{c}}
\newcommand{\baz}{\begin{array}{cc}}
\newcommand{\bad}{\begin{array}{ccc}}
\newcommand{\bea}{\begin{equation} \begin{array}{c}}
\newcommand{\eea}{ \end{array} \end{equation}}
\newcommand{\ea}{\end{array}} 

\newcommand{\dmsol}{\mbox{$\Delta m^2_{\odot}$}}
\newcommand{\dma}{\mbox{$\Delta m^2_{\rm A}$}}

 \def\ra{\rightarrow}

\newcommand{\mmin}{\mbox{$m_0$}}
\def\gtap{\mathrel{ \rlap{\raise 0.511ex \hbox{$>$}}{\lower 0.511ex
   \hbox{$\sim$}}}} 
\def\ltap{\mathrel{ \rlap{\raise 0.511ex
   \hbox{$<$}}{\lower 0.511ex \hbox{$\sim$}}}}
   \newcommand{\deltasol}{\mbox{$ \Delta m^2_{21}$}}
   
   \newcommand{\betabeta}{\mbox{$(\beta \beta)_{0 \nu}$}}

  \newcommand{\meff}{\mbox{$ \left|< \! m \! >
            \right|$}}
  \newcommand{\mefff}{\mbox{$ < \! m \! > $}}

   \newcommand{\hbeta}{$\mbox{}^3 {\rm H}$ $\beta$-decay }
   \newcommand{\eV}{\mbox{$ \ \mathrm{eV}$}}

\newcommand{\pmns}{\mbox{$ U_{\rm PMNS}$}}

\begin{document}

\hfill{Ref.\ SISSA 28/2005/EP} 
 \rightline{CERN-PH-TH/2005-082}
\rightline{May 2005}
\rightline{hep-ph/0505226}

\vspace{0.5cm}

\begin{center}
{\bf\large The Absolute Neutrino Mass Scale, Neutrino Mass Spectrum,\\[2mm]
Majorana CP-Violation and  Neutrinoless Double-Beta Decay}

\vspace{0.5cm}

S. Pascoli~$^{a)}$, \hskip 0.2cm S. T. Petcov~$^{b,c)}$
\footnote{Also at: Institute of Nuclear Research and Nuclear Energy,
Bulgarian Academy of Sciences, 1784 Sofia, Bulgaria} ~and~
T. Schwetz~$^{b)}$

\vspace{0.2cm}

{\em $^{a)}$Theory Division, CERN, CH-1211 Geneva 23, Switzerland\\ }
{\em $^{b)}$Scuola Internazionale Superiore di Studi
Avanzati, I-34014 Trieste, Italy\\ }
{\em $^{c)}$Istituto Nazionale di Fisica Nucleare,
Sezione di Trieste, I-34014 Trieste, Italy\\ }

\end{center}

\vspace{0.5cm}

\renewcommand{\thefootnote}{\arabic{footnote}}
\setcounter{footnote}{0}

\begin{abstract}
Assuming 3-$\nu$ mixing,
massive Majorana neutrinos and
neutrinoless double-beta ($\betabeta$-) 
decay generated only by the $(V\!-\!A)$ 
charged current weak interaction 
via the exchange of the 
three Majorana neutrinos,
we briefly review the predictions 
for the effective Majorana mass 
$\meff$ in $\betabeta$-decay and
reanalyse the physics potential 
of future  $\betabeta$-decay experiments
to provide information on the type 
of neutrino mass spectrum,
the absolute scale of neutrino masses, and 
Majorana CP-violation in the lepton sector. 
Using as input the most recent experimental results
on neutrino oscillation parameters
and the prospective precision that can 
be achieved in future 
measurements of the latter,
we perform a statistical analysis of a 
\betabeta-decay half-life measurement
taking into account experimental and 
theoretical errors, as well as
the uncertainty implied by the 
imprecise knowledge of the 
corresponding nuclear matrix element (NME). 
We show, in particular, how the possibility 
to discriminate between
the different 
types of neutrino mass spectra
and the constraints on the
absolute neutrino mass scale
depend on the 
mean value and the experimental 
error of $\meff$ and on the NME 
uncertainty. The constraints on 
Majorana CP-violation
phases in the neutrino mixing matrix,
which can be obtained 
from a measurement of 
$\meff$ and additional data on the sum 
of neutrino masses,
are also investigated in detail. 
We estimate the required experimental
accuracies on both types of 
measurements, and the required precision
in the NME permitting to address the issue of
Majorana CP-violation in the lepton sector.
\end{abstract}

\newpage
\section{Introduction}

\hskip 1.0truecm   There has been remarkable  
progress in the studies of neutrino
oscillations in the last several years.
The experiments with solar, 
atmospheric, reactor and accelerator neutrinos
\cite{sol,SKsolar,SNO123,SKatm98,SKatmnu04,KamLAND,K2K}
have provided compelling evidence for the existence 
of neutrino oscillations caused by nonzero neutrino 
masses and neutrino mixing
\footnote{Indications for $\nu$-oscillations
were reported also by the 
LSND collaboration \cite{LSND}.
}. 
The latest addition to these 
results are the Super-Kamiokande (SK) 
data on the $L/E$-dependence of 
the (essentially multi-GeV) 
$\mu$-like  atmospheric 
neutrino events \cite{SKdip04},
$L$ and $E$ being the distance traveled 
by neutrinos and the neutrino energy,
and the new spectrum data of the KamLAND 
(KL) and K2K experiments \cite{KL766,K2Knu04}.
For the first time the data directly
exhibit the effects of the 
oscillatory dependence on $L/E$ and $ E$ of 
the probabilities of  
$\nu$-oscillations in vacuum \cite{BP69}.
As a result of these magnificent developments, 
the oscillations of solar $\nu_e$,
atmospheric $\nu_{\mu}$ and $\bar{\nu}_{\mu}$, 
accelerator $\nu_{\mu}$ (at $L\sim$ 250 km)
and reactor $\bar{\nu}_e$ (at $L\sim$ 180 km), 
driven by non-zero $\nu$-masses 
and $\nu$-mixing, can be considered as 
practically established.

  The evidences for $\nu$-oscillations 
obtained in the solar and atmospheric neutrino
and KL and K2K experiments imply  
the existence of 3-$\nu$ mixing
in the weak charged-lepton current:
\begin{equation}
\nu_{l \mathrm{L}}  = \sum_{j=1}^{3} U_{l j} \, \nu_{j \mathrm{L}},~~
l  = e,\mu,\tau,
\label{3numix}
\end{equation}
%
\noindent where 
$\nu_{lL}$ are the flavour neutrino fields,
$\nu_{j \mathrm{L}}$ is the 
field of neutrino $\nu_j$ having 
a mass $m_j$ and
$U$ is the Pontecorvo--Maki--Nakagawa--Sakata (PMNS) 
mixing matrix \cite{BPont57},
$U \equiv \pmns$.
All existing  $\nu$-oscillation 
data, except the data of the LSND experiment~\cite{LSND},
can be described assuming 
3-$\nu$ mixing in vacuum;
we will consider 
this possibility in what follows
\footnote{The interpretation of
LSND data in terms of  
$\bar \nu_{\mu}\to\bar \nu_{e}$ oscillations
requires $(\Delta m^{2})_{\rm{LSND}}\simeq 
1~\rm{eV}^{2}$. The minimal 
4-neutrino mixing scheme, which could 
incorporate the LSND indications for 
$\bar \nu_{\mu}\ra \bar \nu_{e}$ 
oscillations, is strongly disfavored 
by the data \cite{Maltoni4nu}.
The $\nu$-oscillation explanation 
of the LSND results
is possible, assuming 5-neutrino 
mixing \cite{JConrad}.
The LSND results are being tested 
in the MiniBooNE experiment
\cite{MiniB}.}.

The PMNS matrix  can be 
parametrized by 3 angles, 
and, depending on whether the massive 
neutrinos $\nu_j$ are 
Dirac or Majorana particles,
by 1 or 3 CP-violation (CPV) phases 
\cite{BHP80,SchValle80D81}.
In the standardly used parametrization 
(see, e.g.\ \cite{BPP1}),
$\pmns$ has the form:
\bea \label{eq:Upara}
\pmns = \left( \bad 
 c_{12} c_{13} & s_{12} c_{13} & s_{13}  \\[0.2cm] 
 -s_{12} c_{23} - c_{12} s_{23} s_{13} e^{i \delta} 
 & c_{12} c_{23} - s_{12} s_{23} s_{13} e^{i \delta} 
 & s_{23} c_{13} e^{i \delta} \\[0.2cm] 
 s_{12} s_{23} - c_{12} c_{23} s_{13} e^{i \delta} & 
 - c_{12} s_{23} - s_{12} c_{23} s_{13} e^{i \delta} 
 & c_{23} c_{13} e^{i \delta} \\ 
                \ea   \right) 
  {\rm diag}(1, e^{i \frac{\alpha_{21}}{2}}, e^{i \frac{\alpha_{31}}{2}}) \, ,
\eea
%
\noindent where 
$c_{ij} = \cos\theta_{ij}$, 
$s_{ij} = \sin\theta_{ij}$,
the angles $\theta_{ij} = [0,\pi/2]$,
$\delta = [0,2\pi]$ is the Dirac CPV phase and
$\alpha_{21}$, $\alpha_{31}$
are two Majorana CPV phases 
\cite{BHP80,SchValle80D81}. 
One can identify 
the neutrino mass squared difference
responsible for solar neutrino oscillations, 
$\dmsol$, with $\Delta m^2_{21} \equiv m^2_2 - m^2_1$, 
$\dmsol = \Delta m^2_{21} > 0$.
The neutrino mass squared difference
driving the dominant
$\nu_{\mu} \rightarrow \nu_{\tau}$ 
($\bar{\nu}_{\mu} \rightarrow \bar{\nu}_{\tau}$)
oscillations of atmospheric 
$\nu_{\mu}$ ($\bar{\nu}_{\mu}$) 
is then given by
$|\dma|=|\Delta m^2_{31}|\cong |\Delta m^2_{32}|
\gg \Delta m^2_{21}$. 
The corresponding solar and 
atmospheric neutrino mixing angles,
$\theta_{\odot}$ and $\theta_{\rm A}$, 
coincide with $\theta_{12}$ and
$\theta_{23}$, respectively.
The angle $\theta_{13}$ is limited by 
the data from the CHOOZ and Palo Verde
experiments~\cite{CHOOZPV}. 

Thus, the basic phenomenological parameters 
characterising the 3-$\nu$ mixing are:
i) the 3 angles $\theta_{12}$, $\theta_{23}$, $\theta_{13}$,
ii) depending on the nature 
of massive neutrinos 1 Dirac ($\delta$) 
or 1 Dirac + 2 Majorana 
($\delta,\alpha_{21},\alpha_{31}$)
CPV phases, and iii) the 3 neutrino masses, $m_1,~m_2,~m_3$.
Getting  precise information about the 
$\nu$-mixing parameters is 
of fundamental importance for understanding the origin 
of neutrino mixing (see, e.g.~\cite{STPNu04}).

The existing neutrino oscillation data
allow us to determine 
$\dmsol$, $|\dma|$, $\sin^2\theta_{\odot}$ and
$\sin^22\theta_{\rm A}$ with a relatively 
good precision and to obtain 
rather stringent limits on $\sin^2\theta_{13}$ 
(see, e.g.~\cite{SKatmnu04,KL766,BCGPRKL2,3nuGlobal,Maltoni4nu}).
The data imply that $\dmsol = 
\deltasol \sim 8.0\times 10^{-5}~{\rm eV^2}$, 
$|\dma| \sim 2.2\times 10^{-3}~{\rm eV^2}$,
$\sin^2\theta_{\odot} \sim 0.30$,
$\sin^22\theta_{\rm A} \sim 1$ and
$\sin^2\theta_{13} < 0.05$.
The sign of $\dma$, as is well known,
cannot be determined from the
present (SK atmospheric 
neutrino and K2K) data.
In the case of 3-$\nu$ mixing
the two possibilities,
$\Delta m^2_{31(32)} > 0$
or $\Delta m^2_{31(32)} < 0$
correspond to two different
types of $\nu$-mass spectrum:\\
-- {\it with normal hierarchy (or ordering)},
$m_1 < m_2 < m_3$, $\dma=\Delta m^2_{31} >0$, and \\
-- {\it with inverted hierarchy (ordering)}
\footnote{In the
convention we use 
(called A),
the neutrino masses are not 
ordered in magnitude
according to their index number:
$\Delta m^2_{31} < 0$ corresponds to
$m_3 < m_1 < m_2$.
We can also always number the
neutrinos with definite mass, 
in such a way that \cite{BGKP96}
$m_1 < m_2 < m_3$. In this convention
(called B) we have in the case 
of the inverted hierarchy spectrum: 
$\dmsol=\Delta m^2_{32}$, $\dma=\Delta m^2_{31}$.
Convention B is used, e.g.\  in \cite{BPP1,PPSNO2bb}.
},
$m_3 < m_1 < m_2$, $\dma =\Delta m^2_{32}< 0$. \\
\noindent Depending on the sign of \dma, ${\rm sgn}(\dma)$, and 
the value of the lightest neutrino mass,  
${\rm min}(m_j)$, the $\nu$-mass  spectrum can be 
%
\begin{itemize}
\item
{\it Normal Hierarchical (NH)}:\\
$m_1 \ll m_2 \ll m_3$, with
$m_2 \cong \sqrt{\dmsol}\sim 0.009$~eV and 
$m_3 \cong \sqrt{|\dma|}\sim 0.047$~eV;
\item
{\it Inverted Hierarchical (IH)}:\\
$m_3 \ll m_1 < m_2$,
with $m_{1,2} \cong \sqrt{|\dma|} \sim 0.047$~eV;
\item
{\it Quasi-Degenerate (QD)}:\\
$m_1 \cong m_2 \cong m_3$, with 
$m_1 \cong m_2 \cong m_3 \cong m_0$,
$m_j^2 \gg |\dma|$, $m_0 \gtap 0.10$~eV. 
\end{itemize}

The precision on the mixing angles $\theta_{21}$,
$\theta_{23}$, $\theta_{13}$, 
and on $\Delta m^2_{21}$ and $|\Delta m^2_{31}|$, 
can be significantly improved
in future $\nu$-oscillation experiments 
(see, e.g.~\cite{TMU04,Reacth13,shika2,SKGdCP04,BCGPTH1204}).
The sign of $\Delta m^2_{31}$ can be 
determined by studying 
oscillations of neutrinos and
antineutrinos, say, 
$\nu_{\mu} \rightarrow \nu_e$
and $\bar{\nu}_{\mu} \rightarrow \bar{\nu}_e$,
in which matter effects are sufficiently large.
This can be done in long-baseline 
$\nu$-oscillation experiments 
running both with neutrino and antineutrino beams
(see, e.g.~\cite{AMMS99})
or in the neutrino mode only~\cite{LBLnuonly1,LBLnuonly2}.
If $\sin^22\theta_{13}\gtap 0.05$
and $\sin^2\theta_{23}\gtap 0.50$,
information on ${\rm sgn}(\Delta m^2_{31})$
might be obtained in atmospheric neutrino 
experiments by investigating the effects 
of the subdominant transitions
$\nu_{\mu(e)} \rightarrow \nu_{e(\mu)}$
and $\bar{\nu}_{\mu(e)} \rightarrow \bar{\nu}_{e(\mu)}$ 
of atmospheric neutrinos that traverse the 
Earth \cite{JBSP203,Huber:2005ep}.

   The neutrino oscillation experiments, 
however, cannot provide information on the
absolute scale of neutrino masses (or on 
${\rm min}(m_j)$) and thus on the possible hierarchical 
structure (NH, IH, QD, etc.)
of the neutrino mass spectrum. 
The oscillations of flavour neutrinos, 
$\nu_{l} \rightarrow \nu_{l'}$
and $\bar{\nu}_{l} \rightarrow \bar{\nu}_{l'}$,
$l,l'=e,\mu,\tau$, are insensitive to the
nature---Dirac or Majorana---of massive neutrinos 
$\nu_j$; they are insensitive to the Majorana 
CPV phases $\alpha_{21,31}$ \cite{BHP80,Lang87}.
If $\nu_j$ are Majorana fermions,
getting experimental information about the  
Majorana CPV phases in $\pmns$
would be a remarkably challenging problem
\footnote{
The phases $\alpha_{21,31}$
can significantly affect 
the predictions for the 
rates of (LFV) decays
$\mu \rightarrow e + \gamma$,
$\tau \rightarrow \mu + \gamma$, etc.
in a large class of supersymmetric theories
with see-saw mechanism of $\nu$-mass
generation (see, e.g.~\cite{PPY03}).
Majorana CPV phases might be at 
the origin of baryon asymmetry of 
the Universe \cite{LeptoG}.} 
\cite{BargerCP,BGKP96,MajPhase1,PPR1}.
 
  Establishing whether $\nu_j$
are Dirac fermions possessing
distinct antiparticles, 
or are Majorana fermions, i.e.\ spin 1/2 particles that 
are identical with their antiparticles, is 
of fundamental importance 
for understanding the underlying symmetries of 
particle interactions and the origin of 
$\nu$-masses. Let us recall that neutrinos $\nu_j$
with definite mass
will be Dirac fermions if 
particle interactions conserve 
some additive lepton number, e.g.\ the total
lepton charge $L = L_e + L_{\mu} + L_{\tau}$. 
If no lepton charge is conserved, 
the neutrinos $\nu_j$
will be Majorana fermions 
(see, e.g.~\cite{BiPet87}).
The observed patterns of 
$\nu$-mixing and of 
$|\dma|$ and $\Delta m^2_{\odot}$ 
can be related to Majorana $\nu_j$ 
and the existence of 
an {\it approximate} symmetry
corresponding to the conservation of the
lepton charge
$L' = L_e - L_{\mu} - L_{\tau}$
\cite{STP82PD,FPR04}.
The massive neutrinos are 
predicted to be of Majorana nature
by the see-saw mechanism 
of neutrino mass generation \cite{seesaw},
which also provides an  
attractive explanation of the
smallness of neutrino masses 
and, through the leptogenesis theory 
\cite{LeptoG}, of the observed baryon 
asymmetry  of the Universe.
Determining the nature
(Dirac or Majorana)
of massive neutrinos $\nu_j$
is one of the fundamental problems 
in the studies of neutrino mixing
(see, e.g.~\cite{STPNu04}).

  If  neutrinos $\nu_j$ are 
Majorana fermions, processes in which 
the total lepton charge $L$ is not 
conserved and changes by two units, 
such as $K^+ \rightarrow \pi^- + \mu^+ + \mu^+$,
$\mu^+ + (A,Z) \rightarrow (A,Z+2) + \mu^-$, 
etc., should exist
\footnote{The existing experimental 
constraints on the $|\Delta L| = 2$
processes have been discussed 
recently in, e.g.~ \cite{DeltaL2}.}.
The only feasible experiments 
that at present have the potential 
of establishing the Majorana nature 
of massive neutrinos are the
experiments searching for the 
neutrinoless double beta 
($\betabeta$)-decay
$(A,Z) \rightarrow (A,Z+2) + e^- + e^-$
(see, e.g.~\cite{BiPet87,Morales02,APSbb0nu}).
Under the assumption of \betabeta-decay generated
{\it only by the (V-A) charged current 
weak interaction via the exchange of the three
Majorana neutrinos}  $\nu_j$ ($m_j \ltap 1$~eV),
the dependence of the 
\betabeta-decay amplitude $A\betabeta$ 
on the neutrino mass and mixing parameters
factorizes in the  effective Majorana mass 
$\mefff$
(see, e.g.~\cite{BiPet87,ElliotVogel02}):
\begin{equation}
A\betabeta \sim \mefff~\mathcal{M}~,
\label{Abb}
\end{equation}
%
\noindent where $\mathcal{M}$ is the corresponding 
nuclear matrix element (NME) and
\meff\ is given by
\begin{equation}
\meff  = \left| m_1 |U_{\mathrm{e} 1}|^2 
+ m_2 |U_{\mathrm{e} 2}|^2~e^{i\alpha_{21}}
 + m_3 |U_{\mathrm{e} 3}|^2~e^{i\alpha_{31}} \right|~.
\label{effmass2}
\end{equation}
%
\noindent 
If CP-invariance holds
\footnote{We assume that $m_j > 0$ and that
the fields of the 
Majorana neutrinos $\nu_j$ 
satisfy the Majorana condition:
$C(\bar{\nu}_{j})^{T} = \nu_{j},~j=1,2,3$,
where $C$ is the charge conjugation matrix.
}, 
one has \cite{LW81}
$\alpha_{21} = k\pi$, $\alpha_{31} = 
k'\pi$, where $k,k'=0,1,2,...$, and
\begin{equation}
\eta_{21} \equiv e^{i\alpha_{21}} = \pm 1,~~~
\eta_{31} \equiv e^{i\alpha_{31}} = \pm 1 
\label{eta2131}
\end{equation}
%
\noindent represent the relative 
CP-parities of  Majorana neutrinos 
$\nu_1$ and $\nu_2$, and 
$\nu_1$ and $\nu_3$, respectively.
As eq.~(\ref{Abb}) indicates, the observation of  
\betabeta-decay of a given nucleus, and
the measurement of the corresponding
half-life, would allow a determination of
$\meff$ only if the value of the relevant
NME is known with a relatively small uncertainty.

   The experimental searches for $\betabeta$-decay
have a long history 
(see, e.g.~\cite{Morales02,ElliotVogel02}).
The best sensitivity was achieved in the
Heidelberg--Moscow $^{76}$Ge experiment \cite{76Ge00HM}:
\begin{equation}
\meff < (0.35 - 1.05)\ \mathrm{eV},~~~\mbox{at}~90\%~{\rm C.L.},
\label{76Ge00}
\end{equation}
%
\noindent where a factor of 3 uncertainty associated with 
the calculation of the relevant nuclear matrix element 
\cite{ElliotVogel02} is taken into account. 
A similar result has been obtained by 
the IGEX collaboration 
\cite{IGEX00}: $\meff < (0.33$--$ 1.35)$~eV (90\%~C.L.).
A positive signal at 
$> 3\sigma$, corresponding to  
$\meff = (0.1$--$0.9)~{\rm eV}$  at 99.73\% C.L.,
is claimed to be observed in \cite{Klap04}. 
This result will be checked 
in the currently running and future
$\betabeta$-decay experiments. 
Two experiments, NEMO3 (with $^{100}$Mo and 
$^{82}$Se) \cite{NEMO3}
and CUORICINO (with $^{130}$Te) \cite{CUORI},
designed to reach a sensitivity of 
$\meff\sim$(0.2--0.3) eV, 
are taking data. Their first results read (90\% C.L.):
\begin{equation}
\meff < (0.7 \mbox{--} 1.2)~\mathrm{eV}~\mbox{\cite{NEMO3}},\qquad
\meff < (0.2 \mbox{--} 1.1)~\mathrm{eV}~\mbox{\cite{CUORI}},
\label{NEMO3CUOR}
\end{equation}
%
\noindent where the estimated uncertainties 
in the NME are accounted for. 
A number of projects aim 
to reach a sensitivity to 
$\meff\sim$~(0.01--0.05) eV 
\cite{APSbb0nu}: 
CUORE ($^{130}$Te), 
GERDA ($^{76}$Ge),
EXO ($^{136}$Xe), MAJORANA ($^{76}$Ge),
MOON ($^{100}$Mo), XMASS ($^{136}$Xe), 
CANDLES ($^{48}$Ca), SuperNEMO, etc. 
These experiments, in particular, 
can test the positive result 
claimed in \cite{Klap04} and probe the region 
of values of $\meff$ predicted in the case of
IH and QD spectra \cite{PPSNO2bb}.

    In the present article
we reanalyze the potential contribution
that the future planned $\betabeta$-decay 
experiments can make to the studies 
of neutrino mixing. 
The observation of \betabeta-decay
and the measurement of the corresponding 
half-life with a sufficient accuracy,
would not only be a proof that the total 
lepton charge is not conserved in nature 
(see, e.g.~\cite{majorana-nature}), 
but might provide also unique information
i) on the type and possible hierarchical structure
(NH, IH, QD, etc.)
of the neutrino mass spectrum 
\cite{BPP1,PPSNO2bb},
ii) on the absolute scale of 
neutrino masses \cite{PPW}, 
and iii) on the Majorana 
CP-violation phases \cite{BGKP96}.
We consider 3-$\nu$ mixing,
assume massive Majorana 
neutrinos and $\betabeta$-decay 
generated only by the $(V-A)$
charged current weak interaction 
via the exchange of the 
three Majorana neutrinos.
As input in the analysis
we use the results of recent studies of the
precision that can be achieved 
in the measurement of the solar neutrino 
and CHOOZ mixing angles $\theta_{12}$ and $\theta_{13}$, 
and of the neutrino mass squared differences 
$\Delta m^2_{21}$ and $|\Delta m^2_{31}|$,
on which $\meff$ depends.
The uncertainty in the measured value of $\meff$,
which is
due to the imprecise knowledge of
the relevant nuclear matrix elements,
is also taken into account.
All relevant errors are treated in a 
statistically self-consistent manner. 

Our work is a continuation of earlier studies 
\footnote{For an extensive list of references see,
e.g.~\cite{STPFocusNu04}.}  (see,
e.g.~\cite{BPP1,PPSNO2bb,PPR1,PPW,PPRSNO2bb,BGGKP99,WR00,Carlosbb03,AbsNuMassFit}).
It is stimulated by the remarkable progress recently made in the
experimental studies of
$\nu$-oscillations~\cite{SNO123,SKatmnu04,KL766,K2Knu04} and by the
recent analyses
\cite{TMU04,Reacth13,shika2,SKGdCP04,BCGPTH1204,SKGdBV04,TH12} in
which the prospects for high precision determination of
$\sin^2\theta_{12}$, $\sin^2\theta_{13}$, $\Delta m^2_{21}$, and
$|\Delta m^2_{31}|$ in future $\nu$-oscillation experiments have been
extensively investigated. As a result of these studies the experiments
that can provide the most precise measurement of the $\nu$-oscillation
parameters $\meff$ depends on, have been identified and a rather
thorough evaluation of the precision that can be achieved has been
made.  In view of these developments a re-examination of the physics
potential of the future $\betabeta$-decay experiments is both
necessary and timely.

  The outline of the paper is as follows. 
In Section \ref{sec:2} we briefly
discuss the present status of the
determination of, and the prospect 
for improvements of the precision on,
the neutrino oscillation parameters 
relevant to the analysis of 
\betabeta-decay experiments.
In Section \ref{sec:meff} we review 
the predictions for the effective
Majorana mass \meff\ as a function 
of the lightest neutrino mass and the type
of the neutrino mass spectrum, taking into account 
present and prospective uncertainties in 
the neutrino oscillation parameters. In
Section \ref{sec:implications} we present the results of a quantitative
investigation of the potential of a future \betabeta-decay experiment
based on a $\chi^2$-analysis. We show our results as a function of
quantities such as the observed mean value of \meff, its experimental
uncertainty, and the uncertainty in the NME. We evaluate the
possibility to obtain information on the lightest neutrino mass, the
type of $\nu$-mass spectrum and Majorana CPV phases from a
\betabeta-decay experiment. In the latter case 
we take into account the constraint on the sum 
of neutrino masses $\Sigma$, 
which could be provided by cosmological observations, 
and investigate in detail the 
accuracies on \meff\ and $\Sigma$,
required in order to probe
Majorana CP-violation
in the lepton sector. 
Finally we conclude in Section \ref{sec:conclusion}.

\section{The Neutrino Mixing Parameters and $\meff$}
\label{sec:2}

\hskip 1.0truecm  One can express \cite{SPAS94}
the two larger neutrino masses 
in terms of the lightest one,
${\rm min}(m_j) \equiv m_0 \equiv m_\mathrm{MIN}$, 
and of $\dmsol$ and $\dma$~  
\footnote{For a discussion of 
the relevant formalism 
see, e.g.~\cite{BPP1,STPFocusNu04}.
Notice that in~\cite{BPP1}
$m_0$ was used in the case of QD spectrum
to indicate $m_0 \equiv m_1 \simeq m_2 \simeq m_3$.
Here we extend this notation to indicate
the smallest neutrino mass for each type of spectrum.}.
Within the convention we use,
in both cases of 
$\nu$-mass spectrum 
with normal and inverted 
hierarchy, one has:
$\dmsol =\Delta m_{21}^2 > 0$.
For normal hierarchy, ${\rm min}(m_j) = m_1$,
$\dma = \Delta m_{31}^2 > 0$,
$m_2 = (m_1^2 + \dmsol)^{\frac{1}{2}}$,
and $m_3 = (m_1^2 + \dma)^{\frac{1}{2}}$.
If the spectrum is with inverted hierarchy,
${\rm min}(m_j) = m_3$,
$\dma = \Delta m_{32}^2 < 0$ and thus
$m_1 = (m_3^2 + |\dma| - \dmsol)^{\frac{1}{2}} \cong 
(m_3^2 + |\dma|)^{\frac{1}{2}}$,
$m_2 = (m_3^2 + |\dma|)^{\frac{1}{2}}$.
For both types of mass ordering,
the following relations hold:
$|U_{\mathrm{e} 1}|^2 = \cos^2\theta_{\odot} (1 - \sin^2\theta_{13})$, 
$|U_{\mathrm{e} 2}|^2 = \sin^2\theta_{\odot} (1 - \sin^2\theta_{13})$,
and  $|U_{\mathrm{e} 3}|^2 \equiv \sin^2\theta_{13}$,
$\theta_{\odot} \equiv \theta_{12}$.
Thus, in the case 
of interest the effective Majorana mass $\meff$, 
eq.~(\ref{effmass2}), depends in general on:
i) $\dma = \Delta m^2_{31(32)}$,
ii) $\theta_{\odot} = \theta_{12}$ and 
$\Delta m^2_{\odot}= \Delta m^2_{21}$, 
iii) the lightest neutrino mass
$m_0$, iv) the mixing angle $\theta_{13}$, and
v) the Majorana CPV phases $\alpha_{21,31}$. 

    The best fit value and the 
95\% C.L. allowed range of
$|\dma|$ found in a combined analysis of
the atmospheric neutrino 
\footnote{The current
atmospheric neutrino data
are insensitive 
to $\theta_{13}$ satisfying 
the CHOOZ limit 
\cite{SKatmnu04}.}
and K2K data
read \cite{SKatmnu04,Maltoni4nu}: 
\beq 
\label{eq:atmrange}
\ba
|\dma| =2.2\times 10^{-3}~{\rm eV^2}
~, \\  [0.25cm]
|\dma| = (1.7 - 2.9)\times 10^{-3}~{\rm eV^2}~.
\ea
\eeq
%
\noindent 
 Combined 2-$\nu$ oscillation 
analyses of the solar neutrino and  
KL 766.3 tyr spectrum data show 
\cite{KL766,BCGPRKL2} that 
$\dmsol$ and $\theta_{\odot}$ 
lie in the low-LMA region:
$\dmsol = (7.9^{+0.6}_{-0.5})\times 10^{-5}~{\rm eV^2}$, 
$\tan^2 \theta_\odot = (0.40^{+0.09}_{-0.07})$.
The high-LMA solution 
(see, e.g.~\cite{SNO3BCGPR})
is excluded at $\sim 3.3\sigma$.
Maximal solar neutrino mixing
is ruled out at $\sim 6\sigma$;
at 95\% C.L.\ one finds
$\cos 2\theta_\odot \geq 0.28$ \cite{BCGPRKL2},
which has important implications for \meff\ (see further). 
In the case of 3-$\nu$ mixing,
the $\nu_e$ and $\bar{\nu}_e$ survival
probabilities, relevant to the 
interpretation of
the solar neutrino, KL and CHOOZ data, 
depend also on $\theta_{13}$ 
\cite{3nuSP88,STPNu04}. 
A combined 3-$\nu$ oscillation
analysis of these data
gives \cite{BCGPRKL2,Maltoni4nu,3nuGlobal}
\beq
\sin^2\theta_{13} < 0.027~(0.047),~~~~\mbox{at}~95\%~(99.73\%)~{\rm C.L.}
\label{th13}
\eeq
%
Furthermore, such an analysis shows \cite{BCGPRKL2}
that for $\sin^2 \theta_{13} \ltap 0.02$
the allowed ranges of 
$\deltasol$ and $\sin^2\theta_{21}$ do not differ 
substantially from those derived in the 
2-$\nu$ oscillation analyses,
and that the best fit values are practically 
independent of $\sin^2 \theta_{13} < 0.05$. 
The best fit values and the allowed ranges
at 95\%~C.L.\ read
\cite{Maltoni4nu,BCGPRKL2}:
\beq
\label{bfvsol}
\ba
\deltasol = 8.0\times 10^{-5}~{\rm eV^2},~~
\sin^2\theta_{21} = 0.31~, \\[0.25cm]
\deltasol = (7.3 - 8.5) \times 10^{-5}~{\rm eV^2},~~
\sin^2 \theta_{12} = (0.26 - 0.36)~.
\ea
\eeq
%
Existing data allow a determination of
$\dmsol$, $\sin^2\theta_{\odot}$
and $|\dma|$ at 3$\sigma$ with an 
error of approximately 12\%, 24\%, 
and 50\%, respectively. 
These parameters can 
(and very likely will) be measured
with much higher accuracy in the future.
The data from phase-III of 
the SNO experiment \cite{SNO123}
\footnote{During this phase 
the neutral current 
rate will be measured in SNO
with $^3$He proportional counters.}
could lead to a reduction of the error 
in $\sin^2\theta_{12}$ 
to 21\% \cite{SKGdCP04,BCGPTH1204}. 
If instead of 766.3 tyr
one uses simulated 3 ktyr
KamLAND data in the same 
global solar and reactor neutrino 
data analysis, the 3$\sigma$ errors in 
$\Delta m^2_{21}$ and $\sin^2\theta_{12}$ 
diminish to 7\% and 18\% \cite{BCGPTH1204}. 
The most precise measurement of 
$\Delta m^2_{21}$ could be achieved 
\cite{SKGdCP04} using 
Super-Kamiokande doped with 0.1\% of gadolinium 
for detection of reactor 
$\bar{\nu}_e$ \cite{SKGdBV04}:
the SK detector 
gets the same flux of reactor
$\bar{\nu}_e$ as KamLAND and
after 3 years of 
data-taking, $\Delta m^2_{21}$
could be determined with  
an error of 3.5\% at 3$\sigma$ 
\cite{SKGdCP04}. 
A dedicated reactor  
$\bar{\nu}_e$ experiment with a 
baseline $L\sim 60$ km, tuned to the minimum of the
$\bar{\nu}_e$ survival 
probability, 
could provide the most precise 
determination of $\sin^2\theta_{12}$ 
\cite{TH12,BCGPTH1204}:
with statistics of $\sim 60$ GW~ktyr 
and systematic error 
of 2\% (5\%), $\sin^2\theta_{12}$  
could be measured with 
an error of 6\% (9\%) 
at 3$\sigma$ \cite{BCGPTH1204}. 
The inclusion of the uncertainty
in $\theta_{13}$
($\sin^2\theta_{13}<$0.05)
in the analyses increases the 
quoted errors by (1--3)\% to 
approximately 9\% (12\%) 
\cite{BCGPTH1204}.
The highest precision in the determination
of $|\dma| = |\Delta m^2_{31}|$
is expected to be achieved 
from the studies
of $\nu_{\mu}$-oscillations in
the T2K (SK) \cite{T2K} experiment:
if the true $|\Delta m^2_{31}| =
2\times 10^{-3}$~eV$^2$ 
(and true $\sin^2\theta_{23} = 0.5$), 
the 3$\sigma$ uncertainty
in $|\dma|$ is estimated to be 
reduced in this experiment to  
$\sim 12\%$ \cite{TMU04}. The error 
diminishes with increasing 
 $|\Delta m^2_{31}|$.

  In what regards the CHOOZ angle
$\theta_{13}$, there are several 
proposals for reactor $\bar{\nu}_e$ 
experiments with baseline $L\sim$~(1--2) km 
\cite{Reacth13}, which could improve 
the current limit, $\sin^2\theta_{13}<$~0.05, by a 
factor of (5--10): Double-CHOOZ (in France), 
Braidwood (in the USA),
Daya-Bay (USA--China), KASKA (in Japan), etc.
The reactor $\theta_{13}$ 
experiments can compete in sensitivity with
accelerator experiments 
(T2K~\cite{T2K}, NO$\nu$A~\cite{NOvA}) 
(see, e.g.~\cite{TMU04})
and can be done on a relatively short 
(for experiments in this field) time scale.

  Information on the absolute 
scale of neutrino masses
can be derived in \hbeta experiments 
\cite{Fermi34,MoscowH3,MainzKATRIN}
and from cosmological and astrophysical data. 
The most stringent upper 
bounds on the $\bar{\nu}_e$ mass 
were obtained in the Troitzk~\cite{MoscowH3} 
and Mainz~\cite{MainzKATRIN} experiments: 
%
\beq
m_{\bar{\nu}_e}  <  2.3 \eV ~~~\mbox{at}~95\%~\mathrm{C.L.}.
\label{H3beta}
\eeq
%
\noindent We have $m_{\bar{\nu}_e} \cong m_{1,2,3}$
in the case of the QD $\nu$-mass spectrum.
The KATRIN experiment~\cite{MainzKATRIN}
is planned to reach a sensitivity  
of  $m_{\bar{\nu}_e} \sim 0.20$~eV,
i.e.\ it will probe the region of the QD 
spectrum. The Cosmic Microwave Background (CMB) data of the 
WMAP experiment, combined with data from large
scale structure surveys (2dFGRS, SDSS), lead to an 
upper limit on the sum of the neutrino masses\cite{WMAPnu}: 
%
\beq
\sum_{j} m_{j} \equiv \Sigma < (0.7 \mbox{--} 2.0)~{\rm eV~~~\mbox{at}~95\%~C.L.},
\label{WMAP}
\eeq
%
\noindent where we have included a conservative 
estimate of the uncertainty in the upper limit 
(see, e.g.~\cite{Hanne03}).
The WMAP and future PLANCK 
experiments can be sensitive to 
$\Sigma~\cong~ 0.4$~eV.
Data on weak lensing of 
galaxies by large scale structure,
combined with data from the WMAP and PLANCK
experiments, may allow $\Sigma$ to be determined 
 with an uncertainty of 
$\delta \sim (0.04\mbox{--}0.10)$~eV \cite{Hu99}.
Similar sensitivities can be reached by analysing the
distortions in the Cosmic Microwave Background due to
gravitational lensing in a future high sensitivity experiment~\cite{Manoj}.

\section{Predictions for the Effective Majorana Mass $\meff$}
\label{sec:meff}

\hskip 1.0cm  Given $\dmsol$, $|\dma|$, $\theta_{\odot}$ 
and $\sin^2\theta_{13}$, the value of $\meff$ 
depends strongly on the type of the
neutrino mass spectrum (NH, IH, QD, etc.) 
and on  the Majorana CPV phases
of the PMNS matrix, $\alpha_{21,31}$ 
(see eq.\ (\ref{effmass2})).
In what follows we will summarise 
the current status of the predictions
for $\meff$.

\vspace{0.2cm}
{\bf Normal Hierarchical Spectrum.}
In this case $m_1 \ll m_2 \ll m_3$, $m_0 = m_1$, 
and one has \cite{BPP1}
\begin{eqnarray}
\meff &=& \left| \left(m_1\cos^2\theta_\odot  + 
e^{i\alpha_{21}} \sqrt{\dmsol +m_1^2}
\sin^2 \theta_\odot \right)\cos^2\theta_{13}\right. \nonumber\\
&& + \left. 
\sqrt{\dma +m_1^2}\sin^2\theta_{13}~e^{i\alpha_{31}} \right| ,
\label{meffNH1} \\
&\simeq& \left| \sqrt{\dmsol}
\sin^2 \theta_\odot \cos^2\theta_{13} + 
\sqrt{\dma} 
\sin^2\theta_{13} e^{i(\alpha_{31} - \alpha_{21})} \right| ,
\label{meffNH2}
\end{eqnarray}
%
\noindent where we have neglected 
$m_1$ in eq.~(\ref{meffNH2}).
Although one neutrino, $\nu_1$,
effectively ``decouples'' and does not 
contribute to $\meff$,
the value of $\meff$ according to
eq.~(\ref{meffNH2}) still 
depends on the Majorana CPV
phase difference $\alpha_{32} = \alpha_{31} - \alpha_{21}$.
This reflects the fact that in contrast
to the case of massive Dirac 
neutrinos (or quarks),
CP-violation can take place
in the mixing of only two massive 
Majorana neutrinos \cite{BHP80}.   

Since at 95\% (99.73\%) C.L.\ 
we have \cite{Maltoni4nu,BCGPRKL2}
$\sqrt{\dmsol} \ltap 9.2~(9.4)\times 10^{-3}$~eV,
$\sin^2\theta_{\odot} \ltap 0.36~(0.40)$,
$\sqrt{\dma} \ltap 5.4~(5.7)\times 10^{-2}$~eV, 
$\sin^2\theta_{13} <0.027~(0.046)$,
and the largest neutrino mass
enters into the expression for $\meff$ multiplied by
the factor $\sin^2\theta_{13}$,
the predicted value of $\meff$ is 
typically $\sim {\rm few}\times 10^{-3}$~eV: 
for $\sin^2\theta_{13} = 0.03$, 
one finds $\meff \ltap 0.005$~eV (using the data at 95\% C.L.).
Using the best fit values of the indicated 
parameters (see eqs.~(\ref{eq:atmrange}) 
and (\ref{bfvsol})) and the same 
value of $\sin^2\theta_{13} = 0.03$,
we get $\meff \ltap 0.0042$~eV.

   The minimal value of $\meff$ in eq.~(\ref{meffNH2})
is obtained when there is a maximal 
compensation between the 
``solar neutrino'' term, 
$\sqrt{\dmsol} \sin^2 \theta_\odot \cos^2\theta_{13}$,
and the ``atmospheric neutrino'' one, 
$\sqrt{\dma} \sin^2\theta_{13}$.
At 95\% (99.73\%) C.L.\ 
we have $\sqrt{\dma} \sin^2\theta_{13} \ltap
1.5~(2.7)\times 10^{-3}$~eV, while the 
``solar neutrino'' term 
takes values in the interval $(2.1 - 3.2)\times 10^{-3}$~eV    
($(1.9 - 3.6)\times 10^{-3}$~eV). 
Thus, at 95\% C.L.\ the ``solar neutrino'' term
is larger than the ``atmospheric neutrino'' one and 
$\meff$ is bounded from below.
However, this may not be true
considering the current 99.73\%~C.L. 
intervals of allowed values of the 
relevant oscillation parameters. 

   It follows from eq.~(\ref{meffNH1})
and the allowed ranges of values of 
$\dmsol$, $\dma$, $\sin^2\theta_{\odot}$,
$\sin^2\theta_{13}$, as well as of 
the lightest neutrino mass $m_1$ and
CPV phases $\alpha_{21,31}$,  that in the case of 
spectrum with {\it normal hierarchy} 
there can be a complete 
cancellation between 
the three terms in eq.~(\ref{meffNH1}), and
one can have \cite{PPW} $\meff = 0$.

\vspace{0.2cm} 
{\bf Inverted Hierarchical Spectrum.}
For IH neutrino mass spectrum, 
$\dma <0$, $m_0 = m_3$, and
$m_3 \ll m_1 \cong m_2 \cong \sqrt{|\dma|} = \sqrt{\Delta m^2_{23}}$. 
Using eq.~(\ref{effmass2}) we find \cite{BGKP96,BPP1}:
%
\begin{eqnarray}
\meff &\cong& \left|(\cos^2\theta_\odot  + 
e^{i\alpha_{21}} \sin^2 \theta_\odot)\cos^2\theta_{13}
\sqrt{m_3^2 + |\dma|}
+ m_3\sin^2\theta_{13}~e^{i\alpha_{31}} \right| ,
\label{meffIH1} \\
&\cong& \sqrt{m_3^2 + |\dma|} \cos^2\theta_{13} 
\left (1 - \sin^22\theta_{\odot} \sin^2\frac{\alpha_{21}}{2}
\right )^{\frac{1}{2}} ,
\label{meffIH2} \\
&\cong& \sqrt{|\dma|} \cos^2\theta_{13} 
\left ( 1 - \sin^22\theta_{\odot} \sin^2\frac{\alpha_{21}}{2}
\right )^{\frac{1}{2}},
\label{meffIH3}
\end{eqnarray}
%
\noindent where we have neglected 
$m_3\sin^2\theta_{13}$ in eqs.~(\ref{meffIH2}) and 
(\ref{meffIH3}). 
The term $m_3\sin^2\theta_{13}$ can always 
be neglected given the existing data: 
even in the case where the spectrum is 
with {\it partial} inverted hierarchy
and $m_3^2 \sim |\dma|$, the
minimum of the sum of the other 
two terms in $\meff$  satisfies
$\sqrt{m_3^2 + |\dma|}\cos 2\theta_\odot \cos^2\theta_{13}
\gg m_3\sin^2\theta_{13}$, since the data 
on $\theta_\odot$ and  $\theta_{13}$ imply
$ \cos 2\theta_\odot \gg \tan^2\theta_{13}$.
Even though in eqs.~(\ref{meffIH2}) and 
(\ref{meffIH3}) one of the
three massive Majorana neutrinos 
``decouples'', the value of $\meff$
depends on the Majorana CP-violating phase
$\alpha_{21}$. 
It follows from 
eq.~(\ref{meffIH3}) that
%
\begin{equation}
\sqrt{|\dma|}~\cos 2 \theta_\odot~\cos^2\theta_{13}
\leq~ \meff \leq \sqrt{|\dma|}\cos^2\theta_{13}.
\label{meffIH4}
\end{equation}
%
The lower and upper limits correspond
to the CP-conserving cases 
$\alpha_{21} = \pi$ and $\alpha_{21} = 0$. 
Most remarkably, since according to the 
solar neutrino and KamLAND data
$\cos 2 \theta_\odot \sim 0.40$
and $\cos 2 \theta_\odot \gtap 0.28$
at 95\% C.L.,
we get a significant lower limit on $\meff$
exceeding $10^{-2}$~eV
in this case \cite{PPSNO2bb,PPW}.
Using, e.g.\  the best 
fit values of $|\dma|$ and 
$\sin^2\theta_{\odot}$ we find:
$\meff \gtap 0.02$~eV.
The maximal value of $\meff$ is determined 
by $|\dma| $ and, according to
eqs.~(\ref{eq:atmrange}) and (\ref{th13}), 
can reach $\meff \sim 0.055$~eV.
The indicated values of $\meff$ are within 
the range of sensitivity of the next generation of
\betabeta-decay experiments.

  Let us note that if $\dma < 0$, i.e.\ if the
neutrino mass spectrum is with inverted
hierarchy, an upper limit on
$\Sigma = (m_1 + m_2 + m_3) \leq 0.10$~eV
would imply $m_3 \ltap 0.02$~eV
and therefore $m_3^2 \ll |\dma|$. 
In this case the spectrum 
would be of the IH type and 
eqs.~(\ref{meffIH3}) and (\ref{meffIH4})
would be valid. 

   The expression for \meff, 
eq.~(\ref{meffIH3}), permits to relate the value of
$\sin^2 \alpha_{21}/2$ to the experimentally 
measurable quantities \cite{BPP1,BGKP96}
$\meff$, $\dma$ and $\sin^22\theta_{\odot}$: 
%
\begin{equation}
\sin^2  \frac{\alpha_{21}}{2}  \cong
\left( 1 - \frac{\meff^2}{|\dma| \cos^4\theta_{13}} \right) 
\frac{1}{\sin^2 2 \theta_\odot}~.
\end{equation}
%
\noindent
A sufficiently accurate measurement of $\meff$
and of $|\dma|$ and  $\theta_\odot$,
could allow us to get information 
about the value of $\alpha_{21}$. 
If, e.g.\  the data show unambiguously that 
$\meff < \sqrt{|\dma|}\cos^2\theta_{13}$,
that would imply $\alpha_{21}\neq 0$.
If in addition 
the data show that 
$\meff > \sqrt{|\dma|}\cos2\theta_{\odot}\cos^2\theta_{13}$,
one could conclude that $\alpha_{21}$ takes a 
CP-violating value.
 
\vspace{0.2cm} 
{\bf Three Quasi-Degenerate Neutrinos.}
In this case 
$m_0 \equiv m_1 \cong  m_2 \cong m_3$,
$m^2_0 \gg |\dma|$, $m_0 \gtap 0.10$~eV. 
Hence, $\meff$ is essentially independent of
$\dma$ and $\dmsol$, and 
the two possibilities, $\dma > 0$ and 
$\dma < 0$, lead {\it effectively}
to the same predictions for $\meff$.
The mass $m_0$ coincides with the 
$\bar{\nu}_e$ mass $m_{\bar{\nu}_e}$ 
measured in the $^{3}$H $\beta$-decay experiments:
$m_0 = m_{\bar{\nu}_e}$.   
Thus, $m_0 < 2.3 \eV$, or if we use a
conservative cosmological upper limit 
\cite{Hanne03}, $m_0 = \Sigma/3 < 0.7$~eV.  The QD 
$\nu$-mass spectrum is realized for 
values of $m_0$, that can be
measured in the $^3$H $\beta$-decay 
experiment KATRIN~\cite{MainzKATRIN}.
The effective Majorana mass $\meff$ is given by
%
\begin{eqnarray}
\meff &\cong& m_0 \left| (\cos^2 \theta_\odot
 + ~\sin^2 \theta_\odot 
 e^{i \alpha_{21}})~\cos^2\theta_{13} + 
 e^{i \alpha_{31}} \sin^2\theta_{13} \right|, 
\label{meffQD0} \\
&\cong& m_0~\left| \cos^2 \theta_\odot + 
~\sin^2 \theta_\odot e^{i \alpha_{21}} \right|
= m_0~\sqrt{1 - \sin^22\theta_{\odot} 
\sin^2\frac{\alpha_{21}}{2}}.
\label{meffQD1}
\end{eqnarray}
%
Similarly to the case of the IH spectrum, one has:
%
\begin{equation}
m_0~\cos2\theta_\odot  \ltap \meff \leq m_0~. 
\label{meffQD2}
\end{equation}
%
For $\cos 2 \theta_\odot \sim 0.40$,
favored by the 
data, one finds a non-trivial lower limit
on $\meff$, $\meff \gtap 0.08$~eV.
For values of the parameters
allowed at 95\% C.L.\ 
one has $\meff \gtap 0.06$~eV. 
Using the conservative cosmological 
upper bound on 
$\Sigma$ we get $\meff < 0.70$~eV. 
Also in this case one can obtain, 
in principle, direct information
on one CPV phase from the measurement of $\meff$,
$m_0$ and $\sin^2 2 \theta_\odot$:
%
\begin{equation}
\sin^2  \frac{\alpha_{21}}{2}  \cong 
\left( 1 - \frac{\meff^2}{m^2_0} \right) 
\frac{1}{\sin^2 2 \theta_\odot}.
\end{equation}

\begin{figure}[t]
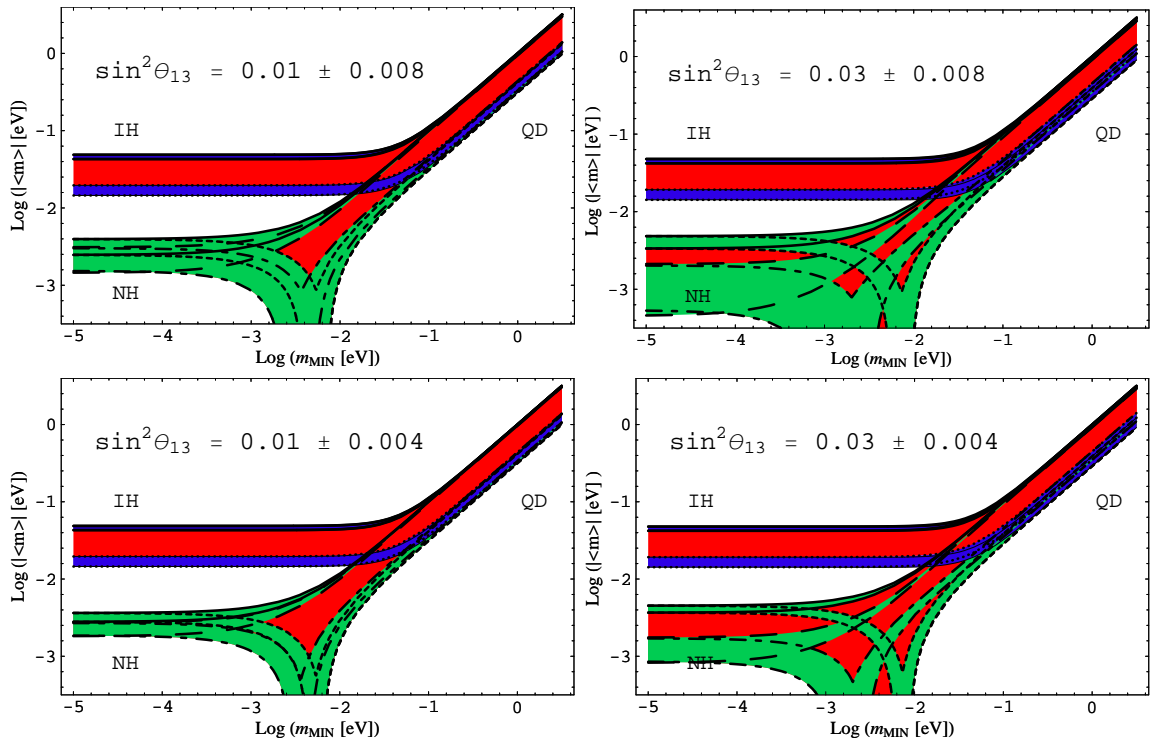

\centering
\vskip -0.3cm
\includegraphics[width=7.5cm]{meff2sigma008_031_01new.epsi}
\includegraphics[width=7.5cm]{meff2sigma008_031_03new.epsi}
\vskip 0.1cm
\includegraphics[width=7.5cm]{meff2sigma004_031_01new.epsi}
\includegraphics[width=7.5cm]{meff2sigma004_031_03new.epsi}
\caption{The predicted value of $\meff$ (including a prospective
   2$\sigma$ uncertainty) as a function of ${\rm min}(m_j)$ for
   $\sin^2\theta_{\odot} = 0.31$ and $\sin^2\theta_{13} = 0.01;~0.03$,
   and two different assumptions on the error on $\sin^2 \theta_{13}$.
   For the NH and QD (and interpolating) spectra, the regions within
   the black lines of a given type (solid, short-dashed, long-dashed,
   dash-dotted) correspond to the four different sets of CP-conserving
   values of the two phases $\alpha_{21}$ and $\alpha_{31}$, and thus
   to the four possible combinations of the relative CP parities
   ($\eta_{21},\eta_{31}$) of neutrinos $\nu_{1,2}$ and $\nu_{1,3}$:
   $(+1,+1)$ solid, $(-1,-1)$ short-dashed, $(+1,-1)$ long-dashed, and
   $(-1,+1)$ dash-dotted lines.  For the IH spectrum, the regions
   delimited by the black solid (dotted) lines correspond to
   $\eta_{21} = + 1$ ($\eta_{21} = -1$), independently of $\eta_{31}$.
   The regions shown in red/medium-gray correspond to violation of
   CP-symmetry (see text for further details).}
\label{Fig1}
\end{figure}
%
  The specific features of the 
predictions for $\meff$ in the cases of 
the three types of neutrino mass 
spectrum discussed above are evident in Fig.~\ref{Fig1}, 
where the dependence of $\meff$ on 
$m_0 = {\rm min}(m_j)$ 
for $\sin^2\theta_{\odot} = 0.31$
and $\sin^2\theta_{13} = 0.01$ and $0.03$
is shown. The figures are obtained
by including a 2$\sigma$ uncertainty
in the predicted value of $\meff$.
The uncertainty in $\meff$, $\sigma(\meff)$,
has been calculated
by exploiting the explicit dependence
of $\meff$ on the oscillation parameters
$\dmsol$, $\dma$, $\sin^2\theta_{\odot}$ and
$\sin^2\theta_{13}$ and assuming 
the following
1$\sigma$ errors (achievable in the future) 
in the determination 
of the latter: $\sigma(\dmsol) = 2\%$, 
$\sigma(|\dma|) =  6\%$,
$\sigma(\sin^2 \theta_{12}) = 4\%$ and
two values of 
$\sigma(\sin^2\theta_{13}) = 0.004$ and $0.008$. 
The current best  fit values of 
$\dmsol$ and $|\dma|$ have 
been used. The Majorana CP-violation phases 
$\alpha_{21}$ and $\alpha_{31}$ 
were varied over all possible 
values they can take
\footnote{It follows from eq.~(\ref{effmass2}) 
that $\meff$ is symmetric under the
transformations $\alpha_{21,31} \to 2\pi - \alpha_{21,31}$. This
implies that it is sufficient to consider values of
$\alpha_{21,31}$ in the range $[0,\pi]$ to cover all possible 
physical configurations for \meff.}.
For the NH and QD (and interpolating) spectra,
the regions within the black lines of a given type  
(solid, short-dashed, long-dashed, dash-dotted)
correspond to the four different sets of
CP-conserving values of the two phases $\alpha_{21}$ and
$\alpha_{31}$, and thus to the four possible combinations of the
relative CP parities ($\eta_{21},\eta_{31}$) of neutrinos
$\nu_{1,2}$ and $\nu_{1,3}$: $(+1,+1)$ solid, $(-1,-1)$ short-dashed,
$(+1,-1)$ long-dashed, and $(-1,+1)$ dash-dotted lines. 
If the spectrum is IH, the contribution to $\meff$ due to $m_3$
can be neglected and the predictions for 
$\meff$ become practically independent of 
$\alpha_{31}$ ($\eta_{31}$).
In this case the regions delimited by the 
black solid (dotted) lines correspond to
$\eta_{21} = + 1$ ($\eta_{21} = -1$).
In the case of CP-violation 
all colored regions are allowed.
The regions shown in red/medium-gray are the so-called
``just CP-violation'' regions~\cite{BPP1}:
an experimental point in these regions
would unambiguously signal CP-violation
associated with Majorana neutrinos. 
In the regions shown in blue/dark-gray and in green/light-gray it is
not possible to distinguish between CP-violation and CP-conservation,
because of the uncertainty implied by the errors on the oscillation
parameters.

   The impact the prospective errors
on $\dma$, $\Delta m_\odot^2$ and $\sin^2 \theta_\odot$
have on the predictions for $\meff$
is, in general, very small.
More specifically, in the case of QD spectrum,
the contributions of
$\sigma(\dma)$,  $\sigma(\Delta m_\odot^2)$ and 
$\sigma(\sin^2 \theta_{13})$ in $\sigma(\meff)$
can be neglected and only 
$\sigma(\sin^2 \theta_\odot)$ induces an
uncertainty in \meff\, which can be as large as
few \% if $\alpha_{21} \sim \pi$.
For the IH type of spectrum,
both $\sigma(|\dma|)$ and $\sigma(\sin^2 \theta_{12})$
are relevant and contribute to an overall 
$\sigma(\meff) \sim {\rm several}$ \%.
Also in this case the error on \meff\
due to $\sigma(\sin^2 \theta_{12})$ increases
as $\alpha_{21}$ varies from 0 to $\pi$.
For NH spectrum, the dominant source of error
is $\sigma(\sin^2 \theta_{13})$.
It affects significantly the predicted value of \meff.
This explains the different 
allowed ranges of values for \meff\ obtained for 
 $\sigma(\sin^2 \theta_{13})=0.004$ and $0.008$.
Notice that the impact of the errors is larger the smaller
$\meff$ is, i.e.\ when $(\alpha_{31}-\alpha_{21})$ approaches
the value $\pi$.
  
  If the spectrum is with normal hierarchy ($\dma >0$),
$\meff$ can lie anywhere
between 0 and the currently existing upper limits, 
eqs.~(\ref{76Ge00}) and (\ref{NEMO3CUOR}).
This conclusion does not change even 
under the most favorable
conditions for the determination of $\meff$,
namely, even when $|\dma|$, $\dmsol$,
$\theta_{\odot}$ and $\theta_{13}$ are known
with negligible uncertainty. 

  The ``gap'' between the predicted values
of $\meff$ in the cases of IH and NH 
spectra allows us, in principle, to
distinguish between these two types of 
{\it hierarchical} spectra \cite{PPSNO2bb,PPW}.  
Establishing, for instance, that
$\meff \neq 0$ but 
$\meff < 10^{-2}$~eV would imply,  
within the 3-neutrino 
mixing scheme with Majorana 
neutrinos under discussion, that
the neutrino mass spectrum is with
normal hierarchy, i.e.\ $\dma >0$. 
Depending on the value of $m_1$, 
the spectrum could be
{\it either normal hierarchical (NH) or with
partial hierarchy} \cite{BPP1}.
Obviously, such a result would rule out 
both the IH and QD spectrum.

   If the results in \cite{Klap04} implying
$\meff = (0.1 \mbox{--} 0.9)~{\rm eV}$ are confirmed,
this would mean, in particular, that
the neutrino mass spectrum is of the QD type.
In this case, however, the measurement of 
$\meff$ cannot provide information on the ${\rm {\rm sgn}}(\dma)$. 
 
   It should be clear from the preceding discussion
that, depending on the measured value of 
$\meff \neq 0$, the $\betabeta$-decay 
experiments may or may not provide
information on {\it both} the type of 
$\nu$ mass spectrum 
(NH, IH, QD, etc.) and  ${\rm sgn}(\dma)$. If 
$\meff \sim {\rm few} \times 10^{-3}~{\rm eV} < 10^{-2}$~eV,
both the type of the spectrum and 
${\rm sgn}(\dma)$ will be determined.
For $\sqrt{|\dma|}~\cos 2 \theta_\odot 
\leq~ \meff \leq \sqrt{|\dma|}$, 
it would be possible to conclude that
${\rm sgn}(\dma) < 0$ only if 
$m_0\ltap 0.02$~eV, i.e. $m_0^2 \ll |\dma|$.
In a relatively narrow interval of values of
$m_0 \sim {\rm few} \times 10^{-2}$~eV, for which 
$m_0^2 \sim |\dma|$,
one can have both $\dma < 0$
and $\dma > 0$. 
In the latter case the $\nu$ mass spectrum is
with {\it partial hierarchy}. If 
$\meff \gtap 0.10$~eV, the $\nu$ mass spectrum
is QD and the measurement of
$\meff$ will provide no information on ${\rm sgn}(\dma)$.

   Finally, if neutrino oscillation experiments show that 
$\dma < 0$ and therefore the $\nu$ mass spectrum 
is with inverted hierarchy, while in $\betabeta$-decay 
experiments only the upper limit
$\meff < \sqrt{|\dma|}\cos2\theta_{\odot}\cos^2\theta_{13}$ is
obtained, that would mean either that there is a new additional 
contribution to the $\betabeta$-decay amplitude 
which interferes destructively with that due to the light 
Majorana neutrino exchange, or that the massive neutrinos 
$\nu_j$ are Dirac particles. Similar conclusion
could be made if, e.g., the KATRIN experiment
shows that $m_0 \gtap 0.2$ eV and correspondingly 
the $\nu$ mass spectrum is QD, while 
$\betabeta$-decay experiments demonstrate only that 
the upper limit $\meff < m_0\cos2\theta_{\odot}$ holds.

\section{Analysis of the Implications of a \betabeta-Decay
Half-Life Measurement}
\label{sec:implications}

\subsection{On the NME Uncertainties}

\hskip 1.0truecm If the \betabeta-decay of a given nucleus is
observed, it will be possible to determine the value of $\meff$ from
the measurement of the associated half-life of the decay.  This would
require the knowledge of the nuclear matrix element of the process.
At present there exist large uncertainties in the calculation of the
\betabeta-decay nuclear matrix elements (see,
e.g.~\cite{ElliotVogel02}). This is reflected, in particular, in the
factor of $\sim 3$ uncertainty in the upper limit on $\meff$, which is
extracted from the experimental lower limits on the \betabeta-decay
half-life of $^{76}$Ge.\footnote{For the uncertainty on the NME for
the \betabeta-decay half-life of $^{76}$Ge commonly a factor of 10 is
adopted. Since \meff\ depends on the square-root of the half-life,
typical values for the current uncertainty on \meff\ are factors from
3 to 4.}  For some nuclei (such as $^{100}$Mo, $^{130}$Te,
$^{136}$Xe), the uncertainties can be even larger.  Recently,
encouraging results on the problem of calculating the nuclear matrix
elements have been obtained in \cite{FesSimVogel03}.  A discussion of
the problems related to the calculation of the $\betabeta$-decay NME
is outside the scope of the present work.  We would like to only note
here that the observation of a \betabeta-decay of one nucleus is
likely to lead to searches and eventually to observation of the decay
of other nuclei.  It can be expected that such a progress will help,
in particular, to solve the problem of the sufficiently precise
calculation of the nuclear matrix elements for the \betabeta-decay
\cite{NMEBiPet04}.

\subsection{The Method of Analysis}
\label{sec:method}

\hskip 1.0truecm The experimental observable in 
\betabeta-decay is the decay rate
$\Gamma_\mathrm{obs}$ measured 
with an experimental accuracy
$\sigma(\Gamma_\mathrm{obs})$. The observed 
decay rate has to be compared with the
theoretically predicted rate
\begin{equation}
\Gamma_\mathrm{th} = G \, |\mathcal{M}|^2 \, \left (\meff
(\boldsymbol{x}) \right )^2\,,
\label{rateth}
\end{equation}
where $G$ is a known 
phase space factor and $\mathcal{M}$ is the NME.
In eq.~(\ref{rateth}) $\boldsymbol{x}=(\boldsymbol{x}_\mathrm{osc},
\boldsymbol{x}^{0\nu}_{\beta\beta})$ are the 
parameters determining $\meff$, which we divide 
into parameters measured in oscillation
experiments, whose values we are going to use as
input in the analysis,
\begin{equation}
\label{eq:osc_params}
\boldsymbol{x}_\mathrm{osc} = (\theta_{12}, \theta_{13}, |\Delta
m^2_{31}|, \Delta m^2_{21})\,,
\end{equation}
%
and parameters that are, in principle, accessible by
\betabeta-decay experiments,
\begin{equation}
\boldsymbol{x}^{0\nu}_{\beta\beta} = 
(\mmin, {\rm sgn}(\Delta m^2_{31}), \alpha_{21},\alpha_{31} )\,.
\end{equation}
%
To investigate the potential to get information on the parameters
$\boldsymbol{x}^{0\nu}_{\beta\beta}$ from the result of a generic
\betabeta-decay experiment, we convert the observed decay rate and the
experimental error into an ``observed effective Majorana mass'' and
its error by
\begin{equation}
\label{eq:convert}
\meff^\mathrm{obs} \equiv 
\sqrt{\frac{\Gamma_\mathrm{obs}}{G}} \, \frac{1}{|\mathcal{M}_0|}\,,\quad
\sigma_{\beta\beta} =
\frac{1}{2} \,
\frac{1}{\sqrt{\Gamma_\mathrm{obs} G}} \, \frac{1}{|\mathcal{M}_0|}\,
\sigma(\Gamma_\mathrm{obs}) \,,
\end{equation}
%
where $|\mathcal{M}_0|$ is some nominal (theoretically predicted)
value of the NME. 
If $\meff^\mathrm{obs} > n\sigma_{\beta\beta}$,
a positive \betabeta-decay signal is observed at the
$n\sigma$~C.L. Otherwise only an upper bound on \meff\ is obtained. 
The quantity $\sigma_{\beta\beta}$ defined in
eq.~(\ref{eq:convert}) is a measure for the ``accuracy'' of the
experiment. Then we construct a $\chi^2$ in the following way:
\begin{equation}\label{eq:chisq}
\chi^2(\boldsymbol{x}^{0\nu}_{\beta\beta},F) =
\min_{\xi \in [1/\sqrt{F}, \sqrt{F}]}
\frac{\left[\xi\, \meff(\boldsymbol{x}) -
\meff^\mathrm{obs}\right]^2} {\sigma_{\beta\beta}^2 +
\xi^2\sigma^2_\mathrm{th}} \,.
\end{equation}
%
\noindent The parameter $\xi$ takes into 
account the uncertainty on
the NME, and it is defined by
\begin{equation}
\xi \equiv \frac{|\mathcal{M}|}{|\mathcal{M}_0|} \,,
\end{equation}
%
where $|\mathcal{M}|$ is the unknown
{\it true} value of the NME and
$|\mathcal{M}_0|$ is the nominal 
value used in eq.~(\ref{eq:convert})
to obtain $\meff^\mathrm{obs}$.  The 
theoretical error $\sigma_\mathrm{th}$ 
in eq.~(\ref{eq:chisq}) takes into account 
the uncertainty implied by the errors on 
the oscillation parameters;
it is calculated by
\begin{equation}
\sigma^2_\mathrm{th} = \sigma^2_\mathrm{th}(\boldsymbol{x}_\mathrm{osc})
=
\sum_i 
\left( \frac{\partial \meff}{\partial x_\mathrm{osc}^i} \right)^2
(\delta x_{\mathrm{osc}}^i)^2 \,,
\end{equation}
%
where the index $i$ runs over the 
four oscillation parameters given in
eq.~(\ref{eq:osc_params}), $\delta x_{\mathrm{osc}}^i$ 
is the uncertainty on the parameter 
$x_{\mathrm{osc}}^i$, and we have
used the fact that to very good approximation
the errors on the oscillation
parameters are uncorrelated 
(see, e.g.~\cite{Maltoni4nu}).

   Assuming that the value of the NME 
is known within a factor $F \ge 1$,
for given parameters $\boldsymbol{x}_{\beta\beta}$ 
we minimise the right-hand side of eq.~(\ref{eq:chisq}) 
with respect to $\xi$, allowing $\xi$ to vary within the 
interval $[1/\sqrt{F},\sqrt{F}]$. A
perfectly known NME corresponds to $F=1$. 
Note that in this way we do not introduce a 
probability weight for the NME; all values between
$|\mathcal{M}_0| /\sqrt{F}$ and $\sqrt{F} |\mathcal{M}_0|$ are treated
on an equal footing.  This procedure is similar 
to the ``flat priors'' used in unitarity triangle 
fits of the CKM matrix in order to account
for theoretical uncertainties, see e.g.~\cite{Ciuchini:2000de}.
We have adopted this method, since it is not possible to assign a well
defined probability distribution to the parameter $\xi$, and
therefore, specifying a range for $\xi$ without imposing any further
weight seems to be the most reliable procedure. Since the choice of
the NME uncertainty factor $F$ is subject to some arbitrariness we
shall show results for various values of $F$.

To combine a measurement of the 
\betabeta-decay rate with a constraint on the 
sum of the neutrino masses $\Sigma$ (obtained, e.g.\ 
from cosmological/astrophysical observations), we generalise
eq.~(\ref{eq:chisq}) in a straightforward way. To take into account
the correlations between $\meff$ and $\Sigma$ induced by the
uncertainties on $\Delta m^2_{21}$ and $\Delta m^2_{31}$, we use the
following covariance matrix in the $\chi^2$-analysis:
\begin{equation}
S_{ab} =
\delta_{ab} (\sigma^\mathrm{exp}_a)^2 +
\sum_i 
\frac{\partial T_a}{\partial x_\mathrm{osc}^i} 
\frac{\partial T_b}{\partial x_\mathrm{osc}^i} 
(\delta x_{\mathrm{osc}}^i)^2 \,,~~~a,b=1,2, 
\end{equation}
%
where $T_1 \equiv \xi \meff$, $T_2 \equiv \Sigma$, and
$\sigma^\mathrm{exp}_{1} \equiv \sigma_{\beta\beta}$ and
$\sigma^\mathrm{exp}_{2} \equiv \sigma_\Sigma$
are the experimental errors on 
$\meff^\mathrm{obs}$ and $\Sigma$, respectively.

\subsection{Constraining the Lightest Neutrino Mass}

\begin{figure}[t]
\centering
\includegraphics[width=0.93\textwidth]{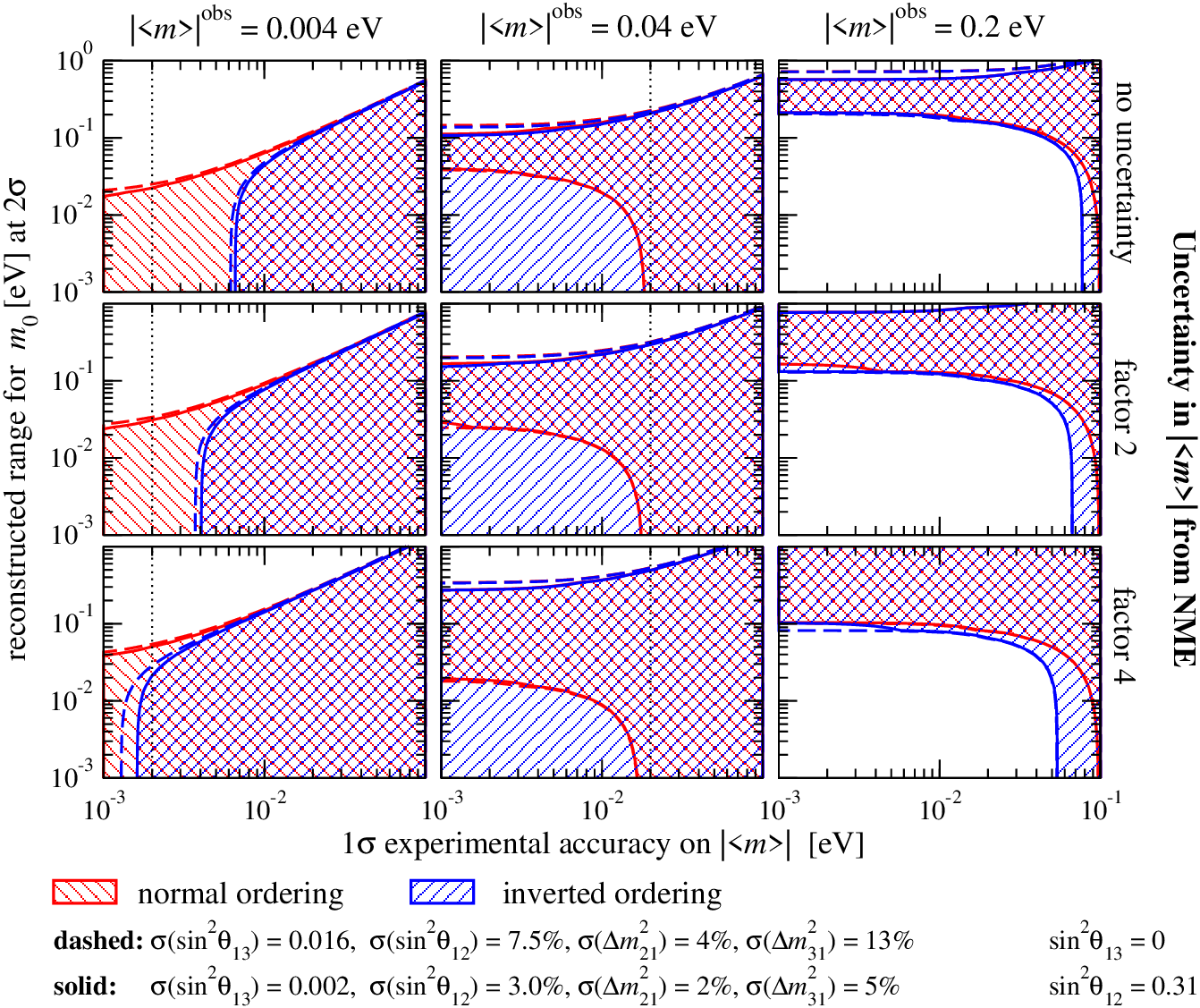}
\caption{The reconstructed range for the lightest neutrino mass
     $\mmin$ at 2$\sigma$~C.L.\ for normal ($\Delta m^2_{31} > 0$) and
     inverted ($\Delta m^2_{31} < 0$) mass ordering as a function of
     the 1$\sigma$ experimental error on $\meff^\mathrm{obs}$. The
     results are shown for three representative values
     $\meff^\mathrm{obs} = 0.004, 0.04, 0.2$~eV (columns of panels),
     and for fixed NME (first row), and an uncertainty of a factor of
     $F=2$ and $F=4$ in the NME (second and third rows).  The figure
     is obtained using the current best fit values of $\Delta
     m^2_{21}$, $|\Delta m^2_{31}|$ and $\sin^2\theta_{12}$
     (eqs.~(\ref{eq:atmrange}) and (\ref{bfvsol})), and
     $\sin^2\theta_{13}=0$.  The dashed (solid) lines correspond to
     the present (prospective) uncertainties on the oscillation
     parameters.  To the left of the dotted lines, a positive signal
     is obtained at $2\sigma$, whereas to the right only an upper
     bound can be stated.}
\label{fig:mlightest}
\end{figure}

\hskip 1.0truecm We start the quantitative evaluation of 
the physics potential of a
$\betabeta$-decay observation by discussing 
the information that can be
obtained on the absolute value of the lightest neutrino mass
$\mmin$. Given an experimental result on $\meff$ from a
$\betabeta$-decay experiment, one can 
infer an allowed range for $\mmin$ for
each type of neutrino mass ordering. 
The results of such an analysis
are shown in Fig.~\ref{fig:mlightest}.  
For given values of $\meff^\mathrm{obs}$ 
and its experimental error $\sigma_{\beta\beta}$,
we minimize the $\chi^2$ of eq.~(\ref{eq:chisq}) 
with respect to the phases $\alpha_{21}$ and 
$\alpha_{31}$, and calculate the allowed
range for $\mmin$ at $2\sigma$ by using the condition 
$\chi^2(\mmin) \le 4$. In Fig.~\ref{fig:mlightest} we adopted 
the best fit values for $\Delta m^2_{21}$, $|\Delta m^2_{31}|$
and $\sin^2\theta_{12}$ (see eqs.~(\ref{eq:atmrange}) and (\ref{bfvsol})), 
and $\sin^2\theta_{13}=0$.
We have verified that the results 
hardly change if the values of the 
oscillation parameters are varied within 
the present $3\sigma$ ranges.  
The dashed lines in 
Fig.~\ref{fig:mlightest}
correspond to the current 
uncertainties of $\Delta m^2_{21}$, 
$|\Delta m^2_{31}|$, $\sin^2\theta_{12}$ 
and $\sin^2\theta_{13}$, while the solid lines
are obtained using 
the following prospective
1$\sigma$ errors:
$\sigma(\Delta m^2_{21}) = 2\%$, 
$\sigma(|\Delta m^2_{31}|) = 5\%$, 
$\sigma(\sin^2\theta_{12}) = 3\%$ 
and $\sigma(\sin^2\theta_{13}) = 0.002$.
By comparing the dashed and solid lines in
Fig.~\ref{fig:mlightest} one observes 
that improving the accuracy of
the oscillation parameters 
has only a minor impact on the 
results: we find only small improvements 
of the constraints on $\mmin$,
while the qualitative 
behavior is unchanged.
%
%
   Consider first the case of 
$ \meff^\mathrm{obs} = 0.2$~eV, shown 
in the right column of Fig.~\ref{fig:mlightest}.
In this case a positive signal 
should be established with
high confidence by the next generation 
of $\betabeta$-experiments. If the
experimental error in $\meff^\mathrm{obs}$
is sufficiently small 
($\sigma_{\beta\beta} \ltap 0.06$~eV for 
NME uncertainty factor $F \leq 3$), 
i) the NH and IH spectra will
be excluded and hence, the neutrino mass spectrum 
will be proved to be QD,
ii) $\mmin$ will be constrained to lie in a rather
narrow interval of values limited from below by $\mmin \gtap 0.1$~eV,
and iii) no information on ${\rm sgn}(\Delta m^2_{31})$ will be
obtained. The uncertainty in the NME directly translates into an
uncertainty in $\mmin$.

In the case of an ``intermediate'' value of $\meff^\mathrm{obs} =
0.04$~eV shown in the middle column of Fig.~\ref{fig:mlightest}, a
lower and an upper bound on $\mmin$ can be established for $\Delta
m^2_{31} > 0$ if $\sigma_{\beta\beta} \ltap 0.017$~eV:
$0.01$~eV~$\ltap \mmin \ltap 0.1$~eV. In the case of $\Delta m^2_{31}
< 0$ only an upper bound will be obtained: $\mmin \ltap 0.1$~eV.  This
result can be easily understood from Fig.~\ref{Fig1}: if $\meff$ is
sufficiently large and $\sigma_{\beta\beta}$ is small enough, the
branch corresponding to the {\it normal hierarchical} 
spectrum extending to $m_0 = 0$ can be excluded.

    Consider finally the left column 
of plots in Fig.~\ref{fig:mlightest}
corresponding to a very small value of
$\meff^\mathrm{obs} = 4\times 10^{-3}$~eV. 
For estimated typical values of $\sigma_{\beta\beta}$ 
of the next generation of $\betabeta$-decay 
experiments and the mean value of
$\meff$ considered, 
only an upper bound on $\meff$ 
can be established.
It is clear that in this case 
one gets also only an upper bound on $\mmin$. 
Moreover, from the panel 
corresponding to a known NME ($F=1$) one
observes that for (ambitious) experimental 
accuracies, i.e. for
$\sigma_{\beta\beta} \ltap 7\times 10^{-3}$~eV,
the case of ${\rm sgn}(\Delta m^2_{31}) < 0$
(inverted mass hierarchy) 
can, in principle, be excluded. 
This is a consequence of the lower bound on 
$\meff$ for inverted ordering, 
which follows from the fact that
$\cos 2\theta_{12}$ is significantly 
different from zero (see eq.~(\ref{meffIH4})
and the related discussion). However, if we take 
into account a possible uncertainty in the NME, 
the requirements on the experimental accuracy
of $\meff$ become exceedingly demanding 
($\sigma_{\beta\beta} \ltap 4\times
10^{-3}$~eV for $F = 2$), 
which renders the exclusion of 
the neutrino mass spectrum 
with inverted hierarchy remarkably 
challenging. Reducing the error to 
$\sigma_{\beta\beta} \cong 10^{-3}$~eV
would allow, e.g.\ for $F\le 2$, 
to conclude that $\mmin \ltap 0.02$~eV and
the neutrino mass spectrum is 
{\it normal hierarchical}.
Establishing in an independent
experiment that ${\rm sgn}(\Delta m^2_{31}) < 0$
would imply in the case under 
consideration that there are additional
mechanism(s) of $\betabeta$-decay~\cite{nonNuMass} 
whose contribution to the $\betabeta$-decay amplitude
compensates partially the one 
due to the Majorana neutrino exchange.

\subsection{Determining the Type of Neutrino Mass Spectrum}

\hskip 1.0truecm As is clear from the previous 
discussions, $\betabeta$-decay experiments provide
a unique possibility to obtain information 
on the type of neutrino
mass spectrum, i.e. to distinguish between 
the NH, IH and QD spectra. 
As we have commented earlier,
getting information on the possible 
hierarchical structure of the neutrino mass 
spectrum and on ${\rm sgn}(\Delta m^2_{31})$
are different, although not totally unrelated,
problems. In this subsection we 
elaborate further on the issue,
since a determination of the 
neutrino mass spectrum is fundamental for
our understanding of neutrino mixing. 
We investigate what conclusions 
can be drawn at the $2\sigma$~C.L.
on the neutrino mass spectrum 
from a result of a $\betabeta$-decay 
experiment, characterised by the observed value 
of $\meff$ and its experimental
error. For given values of $\meff^\mathrm{obs}$,
$\sigma_{\beta\beta}$, the uncertainty 
$F$ in the NME and a fixed neutrino mass ordering, 
we minimize the $\chi^2$-function of
eq.~(\ref{eq:chisq}) with respect to 
$\mmin, \alpha_{21}$ and $\alpha_{31}$. 
If the $\chi^2$-minimum is smaller than 4, 
we conclude that this type of ordering is allowed. 
In addition we test if the ``data'' are consistent with 
negligible $\mmin$, which implies a hierarchical spectrum
(more precisely, we test whether $\chi^2(\mmin=0)\le 4$).
%
\begin{figure}
\centering
\includegraphics[width=0.93\textwidth]{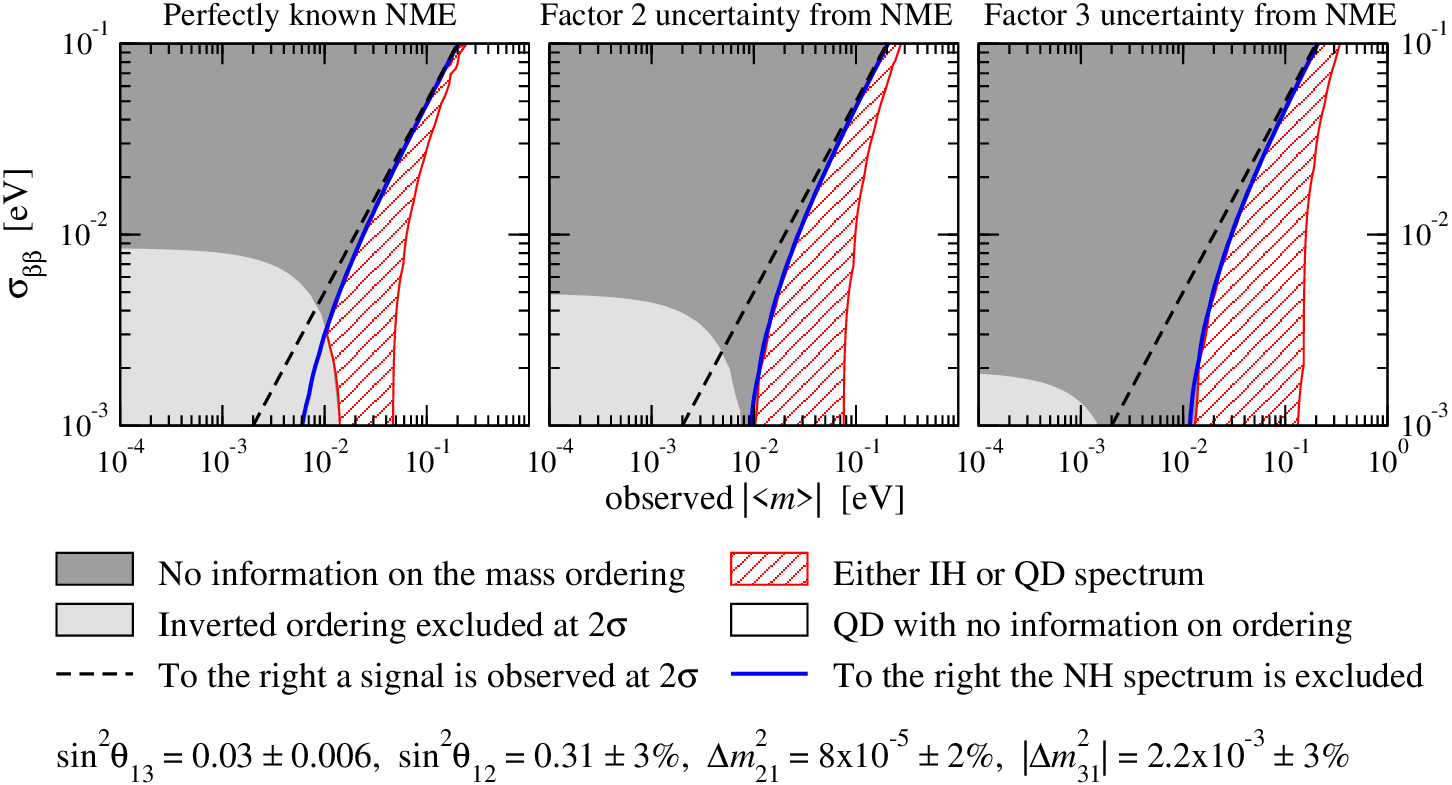}
\caption{Information on the type of neutrino mass spectrum, inferred
   from data on $\betabeta$-decay as a function of the observed
   $\meff$ and its experimental error for three different assumptions
   on the NME uncertainty factor $F$ (see text for details).}
\label{fig:spectrum}
\end{figure}
%
    The results of our analysis are shown
graphically in Fig.~\ref{fig:spectrum}. 
For values of $\meff^\mathrm{obs}$ and
$\sigma_{\beta\beta}$ forming the
dark shaded and white areas in the three panels, no information
on  ${\rm sgn}(\Delta m^2_{31})$ can be obtained.
The light shaded regions correspond to the case where 
${\rm sgn}(\Delta m^2_{31}) < 0 $
(inverted mass ordering) can be excluded. 
In agreement with the results
presented in the previous subsection,
we find that this is only
possible for $\meff^\mathrm{obs} < 0.01$ eV
and an experimental error well below $0.01$~eV.
To the right of the solid curve,
the spectrum cannot be hierarchical
for ${\rm sgn}(\Delta m^2_{31}) > 0$,
i.e.  the possibility
$m_1 \ll m_2 \ll m_3$ is ruled out
(at 2$\sigma$). In the hatched region
in this domain the ``data'' are still 
consistent with $\mmin=0$ for 
the inverted ordering, 
i.e.  with an IH spectrum.
Hence, if a result within 
the hatched region is obtained, we
can conclude that 
either ${\rm sgn}(\Delta m^2_{31}) > 0$ and
the spectrum is with partial hierarchy 
or of the QD type, or 
${\rm sgn}(\Delta m^2_{31}) < 0$ 
and the spectrum is IH 
($m_0 \leq 0.02$ eV), or
QD ($m_0 \gtap 0.1$ eV), or
with partial hierarchy ($m^2_0 \sim  |\dma|$).
This situation corresponds, e.g.\  
to the panels of the middle column in
Fig.~\ref{fig:mlightest}, or to the 
case when the lower branch 
at $\mmin \ltap 0.01$~eV
in the case of normal ordering
(see Fig.~\ref{Fig1}) can be
excluded. Finally, for 
sufficiently large values of 
$\meff^\mathrm{obs}$, corresponding to the
white regions in Fig.~\ref{fig:spectrum}, 
the spectrum is of QD type
and no information on ${\rm sgn}(\Delta m^2_{31})$
can be obtained. 

   Let us add that these results 
are stable with respect to
variations of the oscillation parameters 
within the present allowed
ranges. In particular, they practically do not 
depend on the value of $\sin^2\theta_{13}$:
no significant changes appear 
if, instead of the rather large value 
$\sin^2\theta_{13} = 0.03$ 
adopted in Fig.~\ref{fig:spectrum},
smaller values are used. 
%
\begin{figure}
\centering
\includegraphics[width=0.93\textwidth]{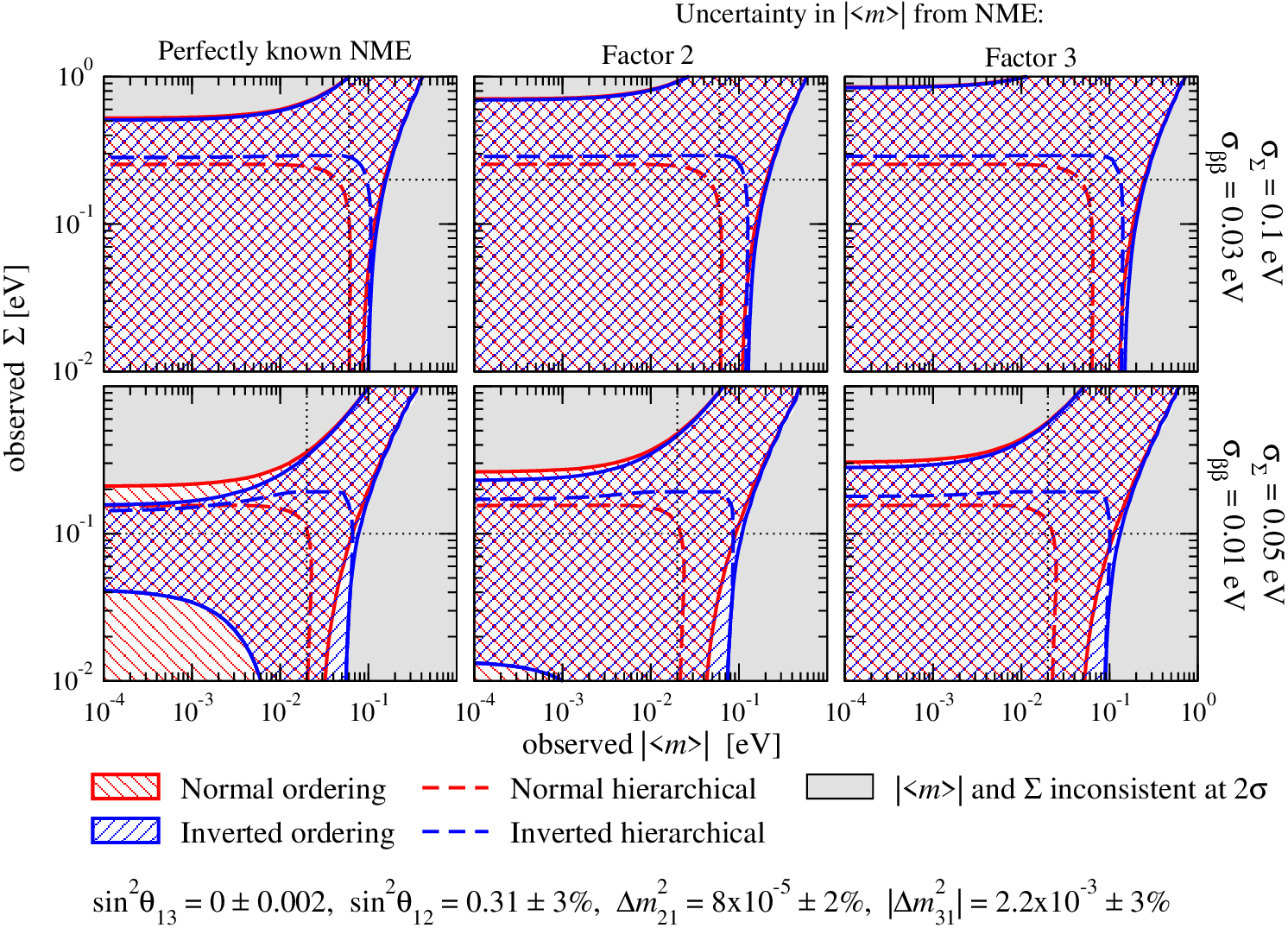}
\caption{Information on the neutrino mass spectrum from a combination
     of \betabeta-decay data on $\meff$ and cosmological data on
     $\Sigma$.  These results are obtained for different assumptions
     about the errors in the determination of $\meff$ and $\Sigma$
     (rows of panels) and the NME uncertainty factor $F$ (columns of
     panels).  We test whether the $\betabeta$-decay and cosmological
     data are consistent with each other, with normal or inverted mass
     ordering, with normal ordering and $\mmin = 0$ (NH $\nu$-mass
     spectrum), or with inverted ordering and $\mmin = 0$ (IH
     spectrum), at $2\sigma$~C.L. The regions to the right of (above)
     the vertical (horizontal) dotted lines correspond to non-zero
     observed $\meff^\mathrm{obs}$ ($\Sigma^\mathrm{obs}$) at
     2$\sigma$.}
\label{fig:cosmo}
\end{figure}
%
As discussed in Section~\ref{sec:2}, 
cosmology provides a sensitive tool
to constrain the sum of the neutrino 
masses $\Sigma \equiv m_1 + m_2 +
m_3$. In the following we investigate 
what can be learned from data on 
$\betabeta$-decay, combined 
with information on $\Sigma$ from cosmology. 
We include the latter by assuming an 
``observed'' value of the sum of the
neutrino masses $\Sigma^\mathrm{obs}$ 
with an experimental accuracy
$\sigma_\Sigma$. Obviously, if 
$\Sigma^\mathrm{obs} > n\sigma_\Sigma$,
cosmological observations 
would provide positive evidence 
for a non-zero $\Sigma^\mathrm{obs}$
at the $n\sigma$~C.L.; otherwise an upper 
bound is obtained.
In Fig.~\ref{fig:cosmo} we 
show the results of a combined 
analysis of a $\betabeta$-decay result with 
information from cosmology as a function of
the observed values of \meff\ and $\Sigma$ 
for two sets of representative experimental 
errors (upper and lower rows) and
different assumptions on the 
uncertainty from the NME. For given
values of $\meff^\mathrm{obs}$ and 
$\Sigma^\mathrm{obs}$ we look for
the $\chi^2$-minimum 
for each of the
two possible mass orderings. 
If a minimum is less than 4, we conclude that 
the corresponding mass ordering is
consistent with the data at $2\sigma$. 
These regions are indicated by
the hatched areas in Fig.~\ref{fig:cosmo}. 
If the $\chi^2$-minimum is
bigger than 4 for both types of mass orderings, 
the corresponding values of $\meff^\mathrm{obs}$ 
and $\Sigma^\mathrm{obs}$ are not
consistent at $2\sigma$ within the assumed 
uncertainties. Such a situation (shown as the 
shaded regions in Fig.~\ref{fig:cosmo})
could either result from systematical 
effects not taken into account
in the cosmological data, or can 
indicate that some mechanism
beyond the light Majorana neutrino exchange is 
operating in \betabeta-decay~\cite{nonNuMass}. 
We also test whether the data are consistent with {\it hierarchical}
spectra, i.e.\ for each sign of $\Delta m^2_{31}$ we test whether
$\chi^2(\mmin=0) \le 4$. These regions are below the dashed lines in
Fig.~\ref{fig:cosmo}, within the corresponding hatched
area.\footnote{We adopt the convention to determine the compatibility
of the \betabeta-measurement and cosmological data by evaluating the
$\chi^2$ for 1 degree-of-freedom. A motivation for this convention is
provided by the so-called parameter-goodness-of-fit method discussed
in~\cite{PG}. In the context of that method the single
degree-of-freedom corresponds to the one parameter, $m_0$, which is
comon to \meff\ and $\Sigma$.}

   For experimental errors corresponding to
$\sigma_{\beta\beta} = 0.03$~eV and
$\sigma_\Sigma = 0.1$~eV adopted in the upper row of plots in
Fig.~\ref{fig:cosmo}, no distinction between 
${\rm sgn}(\Delta m^2_{31}) > 0$ and 
${\rm sgn}(\Delta m^2_{31}) < 0$
(i.e.  normal and inverted ordering) 
is possible. However, some information 
can be obtained on whether the
spectrum is {\it hierarchical} for a given 
${\rm sgn}(\Delta m^2_{31})$.
In particular, the data
from cosmology increase the ability 
to distinguish between IH 
and QD spectra 
in the case of ${\rm sgn}(\Delta m^2_{31}) < 0$
if $\meff \simeq 0.1$~eV is observed. 
This situation corresponds to the
case indicated by the hatched region in 
Fig.~\ref{fig:spectrum}, where
\betabeta-decay alone can only rule 
out the {\it normal hierarchical} spectrum.

   For the more demanding experimental 
precision of $\sigma_{\beta\beta} =
0.01$~eV and $\sigma_\Sigma = 0.05$~eV, 
used in the lower row of plots in 
Fig.~\ref{fig:cosmo},
a new possibility to distinguish 
between normal and inverted mass 
ordering appears. If  the $\betabeta$-decay data 
give, e.g.\ a value of $\meff$ in the interval
(0.04 - 0.07) eV for
NME uncertainty factor $F = 2$, 
and the cosmological observations yield an 
upper bound $\Sigma^\mathrm{obs}
\le 2\sigma_\Sigma = 0.1$~eV, the 
IH spectrum can be established 
at $2\sigma$~C.L. This is a qualitatively 
new method to determine the mass ordering, 
emerging from a synergy between
data from \betabeta-decay experiments 
and cosmology. 
Using $\betabeta$-decay data alone
it is possible, in principle, to rule 
out the inverted ordering if 
$\meff^\mathrm{obs} < 0.01$~eV
and the error $\sigma_{\beta\beta}$
is sufficiently small. 
However, as we have shown, the
required error is exceedingly small:
$\sigma_{\beta\beta} \ltap {\rm few} \times 10^{-3}$~eV. 
In contrast, the conclusion following 
from a combination of \betabeta-decay
and cosmological data is based on 
i) the observation of a value of $\meff$ 
compatible with those predicted 
for the IH spectrum, and ii) an upper
bound on $\mmin$ from cosmological data 
such that the region where the 
predictions for $\meff$ in the cases 
of normal and inverted hierarchies merge, 
can be excluded (see Fig.~\ref{Fig1}). 
Note that this possibility  
remains even if a factor 3 
uncertainty in the NME is taken into account.

\subsection{Constraining the Majorana CPV Phases}
\label{sec:CPV}

\hskip 1.0truecm The possibility of establishing CP-violation due to
Majorana phases through the observation of \betabeta-decay has been
studied previously in~\cite{PPW,PPR1,BargerCP}. In the following we
re-consider this problem by applying the $\chi^2$-method described in
Section~\ref{sec:method}.  We discuss first the 
case of QD spectrum and later consider the possibility
of spectrum with inverted hierarchy, ${\rm sgn}(\dma) < 0$,
and $\meff^\mathrm{obs}$ lying in the IH range of $(0.015 - 0.05)$~eV.
As is clear from eqs.~(\ref{meffIH1})
and (\ref{meffQD0}) and the discussion in Section~\ref{sec:meff}, the
dependence of $\meff$ on the phase $\alpha_{31}$ is suppressed by the
small value of $\sin^2\theta_{13}$. Therefore, we concentrate on the
determination of the phase $\alpha_{21}$, while the dependence on
$\alpha_{31}$ is taken into account implicitly by minimising the
$\chi^2$ with respect to it. For the oscillation parameters we will
assume throughout this subsection uncertainties at the few percent
level (precise numbers are given in the figures).  Such a precision
can be reached in upcoming oscillation experiments.
%
\begin{figure}
\centering
\includegraphics[width=0.93\textwidth]{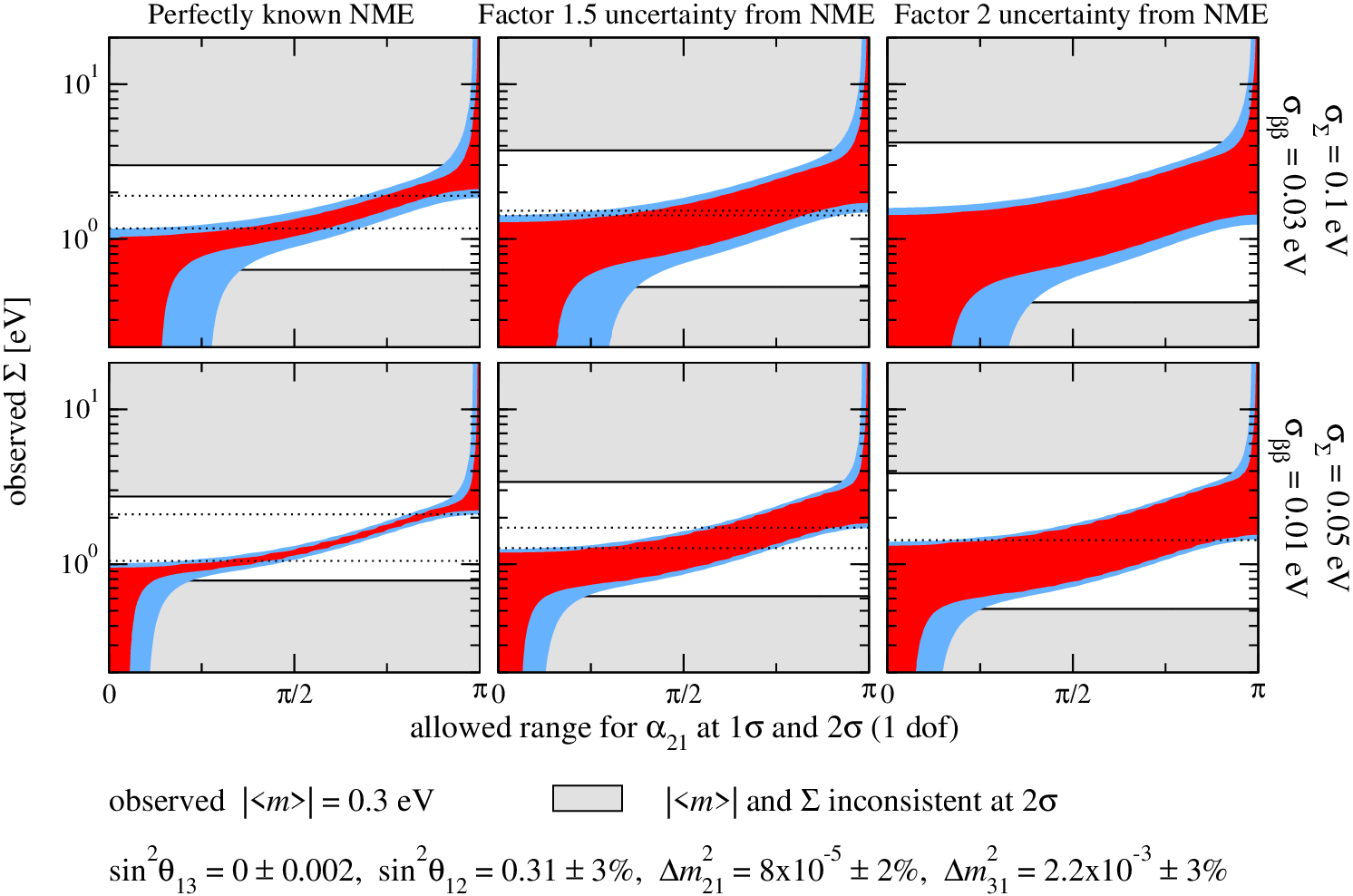}
\caption{Allowed range for the Majorana phase $\alpha_{21}$ at
   1~$\sigma$~C.L.\ (dark-gray/red regions) and 2~$\sigma$~C.L.\
   (medium-gray/blue regions) for $\meff = 0.3$~eV as a function of
   the observed value of $\Sigma$. The shown results are obtained for
   two sets of assumed errors in the observed $\meff$ and $\Sigma$
   (rows of panels) and three values of the NME uncertainty factor $F$
   (columns of panels).  For values of the parameters in the regions
   between the dotted lines, Majorana CP-violation can be established
   at $2\sigma$.}
\label{fig:alpha}
\end{figure}
%
In Fig.~\ref{fig:alpha} we show 
the allowed range for the Majorana
phase $\alpha_{21}$ for $\meff^\mathrm{obs} = 0.3$~eV 
as a function of the observed mean value of $\Sigma$. 
The $1\sigma$ ($2\sigma$) range is
obtained by the condition $\Delta\chi^2(\alpha_{21}) =
\chi^2(\alpha_{21}) - \chi^2_\mathrm{min} \le 1\,(4)$. Since the
allowed range is determined by $\Delta\chi^2$ with respect to the
$\chi^2$-minimum, there is always an ``allowed region'',
irrespectively of whether $\Sigma^\mathrm{obs}$ 
is consistent with the adopted value of
$\meff^\mathrm{obs}$. 
We indicate in Fig.~\ref{fig:alpha} 
the region where the ``results'' of \betabeta-decay
experiment and cosmological observations
are inconsistent ($\chi^2_\mathrm{min} \ge 4$) by the
light shading. Majorana CP-violation can be 
established if both $\alpha_{21} =0$ 
and $\alpha_{21} =\pi$ can be excluded.
The relevant regions are 
indicated by the horizontal dotted lines 
in Fig.~\ref{fig:alpha}. One observes that for
$\sigma_{\beta\beta} = 0.03$~eV and $\sigma_\Sigma =
0.1$~eV used in the upper row of plots, 
already an uncertainty of a
factor of 1.5 in the NME makes it practically 
impossible to establish CPV. 
Our results show, in agreement with 
the results of the previous studies~\cite{PPW,PPR1}, 
that establishing Majorana CP-violation 
due to $\alpha_{21}$ is very challenging: 
the errors in the observed 
$\meff$ and $\Sigma$ 
should not exceed approximately 10\% 
and the NME has to be known within a 
factor $F \ltap 1.5$.
Although establishing Majorana CPV would be a 
very difficult task, it could be possible 
to exclude a certain fraction of the 
full parameter space of the phase $\alpha_{21}$
\footnote{This situation is similar to
the case of the determination of the Dirac CP phase $\delta$ by
long-baseline oscillation experiments, see~\cite{Huber:2004gg} for a
recent discussion and a list of references.}
by using the data on $\meff$ and $\Sigma$. 
In particular, in many cases
it could be possible to exclude one of the 
CP-conserving values of $\alpha_{21}$,
$\alpha_{21} =0$ or $\alpha_{21} =\pi$, 
corresponding to specific
relative CP-parities of the neutrinos 
$\nu_1$ and $\nu_2$.
%
\begin{figure}[t]
\centering
\includegraphics[width=0.93\textwidth]{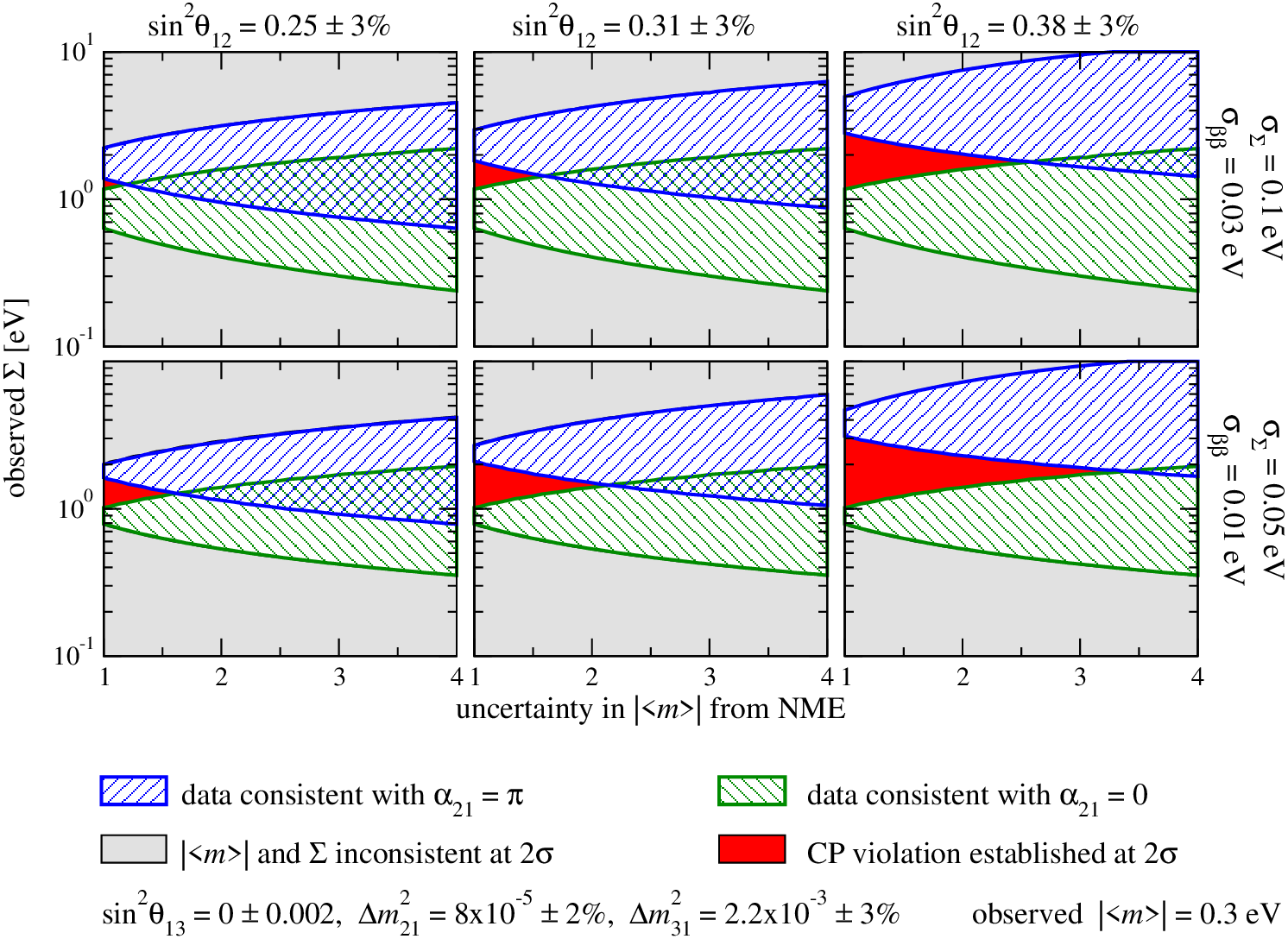}
\caption{Constraints on the Majorana phase $\alpha_{12}$ at 95\%~C.L.\
   from an observed $\meff^\mathrm{obs} = 0.3$~eV and (cosmological)
   ``data'' on $\Sigma$, as a function of the NME uncertainty factor $F$.
   Shown are the regions in which i) the ``data'' are consistent with
   one of the CP-conserving values $\alpha_{12} = 0$ or $\pi$
   (hatched), ii) $\Sigma^\mathrm{obs}$ is inconsistent with
   $\meff^\mathrm{obs} = 0.3$~eV (light-shaded), and iii) Majorana
   CP-violation is established (red/dark-shaded).  The results are
   presented for three values of $\sin^2\theta_\odot$ within the
   currently allowed range (columns of panels), and for two choices of
   the experimental accuracies for \meff\ and $\Sigma$ (rows of
   panels).}
\label{fig:CPV}
\end{figure}
%
   The sensitivity to $\alpha_{21}$ depends 
significantly on the value of
the mixing angle $\theta_\odot$ \cite{PPW,PPR1}. 
As is discussed in Sec.~\ref{sec:meff}
(see eq.~(\ref{meffQD2})), for fixed \mmin\ the allowed range of
\meff\ is given by $\mmin\cos 2\theta_\odot \le \meff \le \mmin$.
Therefore, the allowed range increases for smaller 
values of $\cos 2 \theta_\odot$, 
which makes it easier to exclude the extreme values of
$\meff$, corresponding to the 
CP-conserving configurations. This
effect is clearly shown in Fig.~\ref{fig:CPV}, 
where the three columns of
panels correspond to different values 
of $\sin^2\theta_\odot$. We use
the current best fit point as well as 
values between the present $2\sigma$ and 
$3\sigma$ limits.  In Fig.~\ref{fig:CPV} we assume an 
observation of $\meff^\mathrm{obs}
= 0.3$~eV, which implies a QD spectrum. 
Adopting representative
values for the experimental accuracies 
on $\meff$ and $\Sigma$, and
scanning values of $\Sigma^\mathrm{obs}$ and 
the uncertainty in the NME, 
we test first whether $\Sigma^\mathrm{obs}$ and
$\meff^\mathrm{obs}$ are consistent 
at 2$\sigma$ C.L. ($\chi^2_\mathrm{min} \le 4$). 
If they are, we test whether the phase $\alpha_{21}$ is
consistent with the CP-conserving values 0 and $\pi$
($\chi^2(\alpha_{21} = 0,\pi) \le 4$). 
These two cases are marked
by the hatched regions in Fig.~\ref{fig:CPV}. 
The double-hatched areas, where the regions 
for $\alpha_{21} = 0$ and $\pi$ 
overlap, correspond to the worst situation, 
where no information on $\alpha_{21}$ can be
obtained and the full range $[0,\pi]$ is allowed by the data.
 If the values of $\Sigma^\mathrm{obs}$ and $\meff^\mathrm{obs}$ are
consistent and both CP-conserving solutions for $\alpha_{21}$ can be
excluded, Majorana CP-violation can be established 
at 95\%~C.L., and we indicate the
corresponding regions in red/dark-shading.
Clearly, establishing Majorana CP-violation 
becomes possible only under rather specific conditions. 
For $\sin^2\theta_\odot \gtap 0.31$ 
and $\sim 10\%$ errors in the measured
$\meff^\mathrm{obs}$ and $\Sigma^\mathrm{obs}$
(upper middle and right panels in Fig.~\ref{fig:CPV}),
the NME has to been known to better 
than within a factor of 1.5.
For smaller values of the errors,
$\sigma_{\beta\beta} \sim 0.01$~eV
and $\sigma_\Sigma \sim 0.05$~eV,
Majorana CP-violation could be 
established even for $F \cong 2$
(lower middle and right panels).
If, however, $\sin^2\theta_\odot \cong 0.25$,
the NME uncertainty has to be 
small, $F \leq 1.5$ and the indicated 
high precision in the measurement of 
$\meff^\mathrm{obs}$ and $\Sigma^\mathrm{obs}$
has to be achieved. Finally, the
Majorana phase $\alpha_{21}$ 
has to have a value approximately 
in the interval $\sim (\pi/4 - 3\pi/4)$.
%
\begin{figure}[t]
\centering
\includegraphics[width=0.93\textwidth]{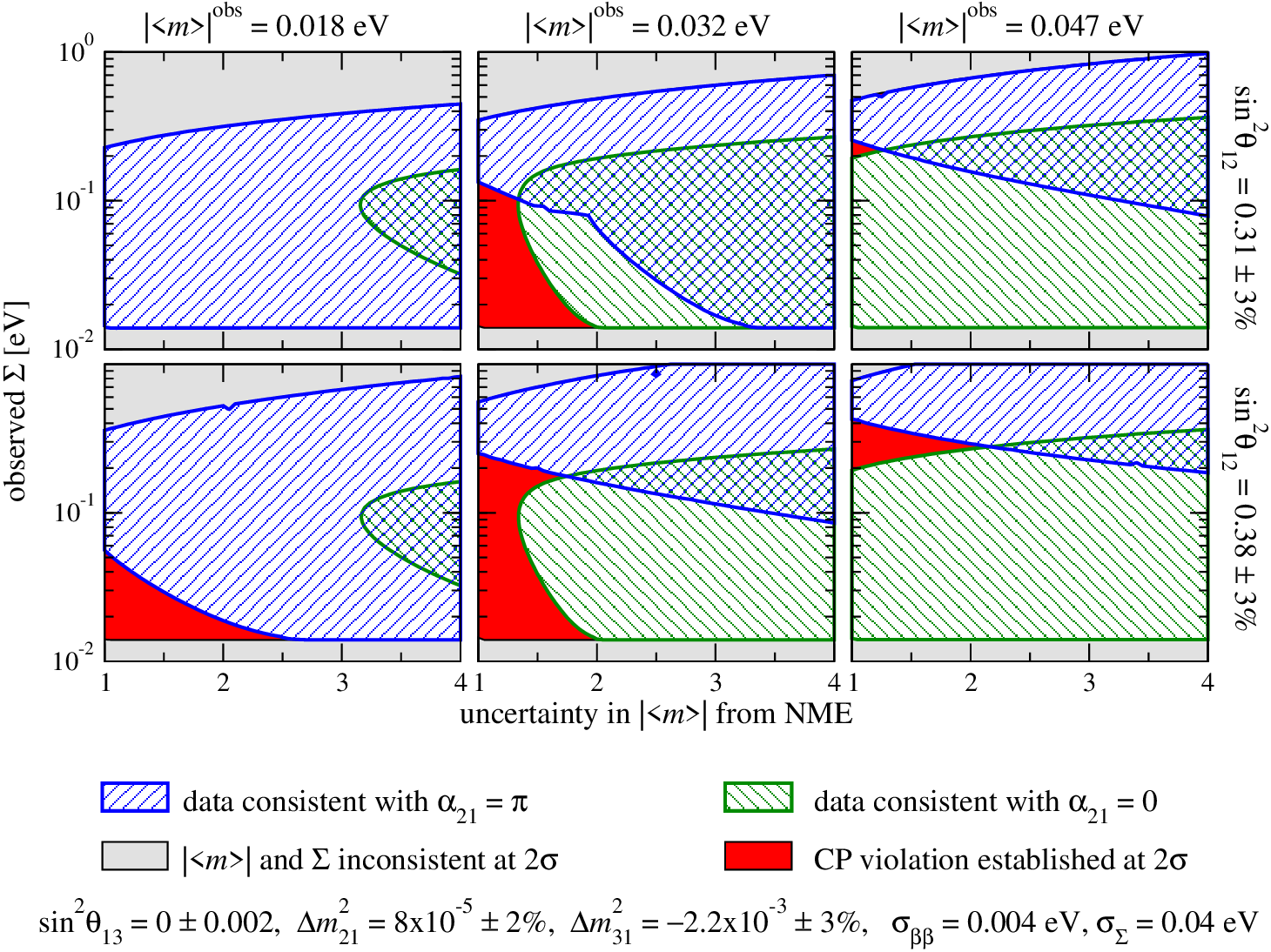}
\caption{Constraints on the Majorana phase $\alpha_{12}$ at 95\%~C.L.\
   for the inverted mass ordering from observed values
   $\meff^\mathrm{obs} = 0.018,\, 0.032, \, 0.047$~eV and
   (cosmological) ``data'' on $\Sigma$, as a function of the
   NME uncertainty factor $F$.  Shown are the regions in which i) the
   ``data'' are consistent with one of the CP-conserving values
   $\alpha_{12} = 0$ or $\pi$ (hatched), ii) $\Sigma^\mathrm{obs}$ is
   inconsistent with $\meff^\mathrm{obs}$ (light-shaded), and iii)
   Majorana CP-violation is established (red/dark-shaded). The upper
   (lower) row of panels corresponds to $\sin^2\theta_\odot =
   0.31;~(0.38)$.}
\label{fig:CPV-IH}
\end{figure}

  Consider next the possibility to establish Majorana 
CP-violation  assuming that the $\nu$ mass spectrum is 
known to be with inverted ordering,
$sgn(\dma) < 0$, and  that
the observed value of $\meff$ lies in the IH region of 
a $\mathrm{few} \times 10^{-2}$~eV. 
Knowing that this would require a rather precise 
measurement of \meff, we use
for the experimental error on \meff\ the value
$\sigma_{\beta\beta} = 4\times 10^{-3}$~eV.
For the sum of neutrino masses $\Sigma$ we adopt
the error $\sigma_\Sigma = 4\times 10^{-2}$~eV.
In Fig.~\ref{fig:CPV-IH} we show 
the sensitivity to Majorana CP-violation for three
representative mean values of \meff\ from the IH region, 
$\meff^\mathrm{obs} = 0.018;~0.032;~0.047$ eV,
and two mean values of $\sin^2\theta_\odot$,
$\sin^2\theta_\odot = 0.31;~0.38$.
The allowed regions in all panels of this figure
are bounded from below
by a straight line at $\Sigma^\mathrm{obs} = 0.014$~eV: 
below this value $\Sigma$ becomes inconsistent 
with the adopted value of $|\dma|$.

   The upper row of panels corresponds to 
the present best fit point of
$\sin^2\theta_\odot$. For $\meff^\mathrm{obs} = 0.018$~eV (left panel), 
the $2\sigma$ interval of allowed values of 
\meff\ {\it always} includes the 
minimal value of \meff\ for the IH
spectrum (see eq.~(\ref{meffIH4})): $\meff^\mathrm{min} \cong \sqrt{|\Delta
m^2_A|}\cos2\theta_\odot$. The latter corresponds to 
the CP-conserving value $\alpha_{21} =\pi$. Thus, 
for the chosen values of $\meff^\mathrm{obs}$,
$\sigma_{\beta\beta}$, $\sigma_\Sigma$
and $\sin^2\theta_\odot$, it is impossible to 
establish Majorana CP-violation.
However, if $F \le 3$, it will be possible to conclude that
$\alpha_{21}$ has a nonzero value, $\alpha_{21} \neq 0$. 
Thus, if CP is conserved, the neutrinos $\nu_1$ and $\nu_2$ 
cannot have the same CP-parities.

    The value $\meff^\mathrm{obs} = 0.032$~eV adopted 
in the middle panel corresponds  
to the case when \meff\ (with the experimental 
uncertainty included) satisfies 
$\meff^\mathrm{min} < \meff < \meff^\mathrm{max}$,
$\meff^\mathrm{max}$
being  maximal value 
of \meff\ predicted in the case of IH spectrum
(see eq.~(\ref{meffIH4})),
$\meff^\mathrm{max} \cong \sqrt{|\Delta m^2_A|}$. 
If an upper bound on \mmin\ is provided by 
a constraint on $\Sigma$,  Majorana CP-violation  can be
established, as evident from the red/dark-shaded 
region in the panel. If the observed value of 
$\Sigma$ becomes too large, the ``data'' becomes 
consistent with the ``upturn'' 
of the IH-branch (see Fig.~\ref{Fig1}), which implies that 
the CP-conserving value $\alpha_{21} = \pi$ is allowed. 
For $\Sigma^\mathrm{obs} \ltap 0.2$~eV
and uncertainties in the NME $F \gtap 1.5 - 2$, 
the ``data'' become consistent with 
$\meff^\mathrm{max}$,
i.e.\ with $\alpha_{21} = 0$.

  For the third representative value of 
 $\meff^\mathrm{obs} = 0.047$~eV (right panels), 
$\meff^\mathrm{max}$ lies in 
the $2\sigma$ interval of allowed values of 
\meff. This implies that Majorana CP-violation
can only be established if the ``data'' on $\Sigma$
constrains \mmin\ precisely in the range 
corresponding to a neutrino mass spectrum with 
{\it partial inverted hierarchy}, 
such that neither $\meff^\mathrm{max}$ 
(i.e.\ the horizontal branch at $\alpha_{21}=0$), nor the 
upturn of the lower (minimal) branch at
$\alpha_{21}=\pi$, are compatible with 
the ``data''. As can be seen in
Fig.~\ref{fig:CPV-IH}, this is marginally possible for
$\sin^2\theta_\odot = 0.31$ (upper row), but some window exists for
$\sin^2\theta_\odot = 0.38$ (lower row) if the NME uncertainty
factor $F \le 2$.

\section{Conclusions} 
\label{sec:conclusion}

\hskip 1.0truecm  In the present article 
we have reanalysed
the potential contribution of future 
$\betabeta$-decay experiments to the studies of
neutrino mixing. We have considered 3-$\nu$ mixing and assumed massive Majorana
neutrinos and $\betabeta$-decay generated only by the $(V-A)$ charged
current weak interaction via the exchange of the three Majorana
neutrinos. In this framework we 
investigated which information can be
obtained from a measurement of the 
effective Majorana mass \meff\  
i) on the type of neutrino mass spectrum (NH, IH, QD, etc.)  ii) on
the absolute scale of neutrino masses, and iii) on the Majorana
CP-violating phases. As input in the analysis 
we used the results of recent studies 
of the prospective precision 
that can be achieved in   
the future measurements of neutrino 
oscillation parameters
on which $\meff$ depends. We performed a $\chi^2$ analysis taking into
account experimental and theoretical errors, as well as the
uncertainty implied by the imprecise knowledge of the corresponding
nuclear matrix element (NME).

   We show how the possibility to discriminate between
the NH, IH and QD spectra depends on the mean value and the
experimental error of $\meff$, and on the NME
uncertainty. Furthermore, we combine the information on \meff\ from a
\betabeta-decay experiment, with a constraint on the sum of the
neutrino masses, $\Sigma$, which can be obtained from cosmological
observations. In this case, we investigate the role of 
the accuracies on \meff\ and
$\Sigma$, as well as on the NME uncertainty, 
in determining the type of neutrino mass spectrum.  
The constraints on Majorana CP-violation phases 
in the neutrino mixing matrix, that can
be obtained from a measurement of $\meff$ and $\Sigma$
in the cases when i) the observed
$\meff \sim {\rm few}\times 10^{-1}$~eV (QD spectrum), and
ii) ${\rm sgn}(\dma) < 0$ and the 
observed $\meff \sim {\rm few}\times 10^{-2}$~eV, 
are also analyzed in detail. We have estimated 
the required experimental accuracies on both 
types of measurements, and the required precision
in the NME permitting to address the issue 
of Majorana CP-violation in the lepton sector.

    Our results show that, in general, getting
quantitative information on the neutrino 
mass and mixing parameters from a measurement of 
the $\betabeta$-decay half-life is 
rather insensitive to the errors on the 
input neutrino oscillation parameters
as long as the errors are smaller than $\sim10\%$.
However, constraints on the absolute neutrino 
mass scale, on the type of neutrino 
mass spectrum and on the Majorana  CP-violation phase 
one can obtain depend critically
on the measured mean value of
$\meff$ (and $\Sigma$), on the precision
reached in the measurement of $\meff$ (and $\Sigma$),
and on the uncertainty in the knowledge of the value of 
the relevant $\betabeta$-decay nuclear matrix element.
The most challenging of these physics 
goals is obtaining quantitative  
information on Majorana CP-violation phases. 
The sensitivity to the latter depends crucially
{\it also} on the value of $\sin^2\theta_{\odot}$.
Establishing Majorana CP-violation using 
data on $\meff$ and $\Sigma$ in the case 
of QD spectrum, for instance,  would require 
for $\sin^2\theta_{\odot} \cong 0.31$, 
a $\sim 10\%$ (or smaller) errors in the measured
$\meff^\mathrm{obs}$ and $\Sigma^\mathrm{obs}$
and knowledge of the relevant
NME with an uncertainty corresponding 
to a factor $F \leq 1.5$.
For smaller values of the errors
and/or larger values of $\sin^2\theta_{\odot}$,
say $\sin^2\theta_{\odot} \cong 0.38$,
it could be possible to obtain evidence of
Majorana CP-violation at 2$\sigma$ C.L. 
even for $F \cong 2$. If, however, 
$\sin^2\theta_\odot \cong 0.25$,
exceedingly high precision in the measurements of 
$\meff$ and $\Sigma$ and NME uncertainty smaller
than 1.5 is required. In all these cases 
the Majorana phase $\alpha_{21}$ 
has to have also a value approximately 
in the interval $\sim (\pi/4 - 3\pi/4)$.

  Future $\betabeta$-decay experiments 
have a remarkable physics potential. They can 
establish the Majorana nature of 
neutrinos with definite mass $\nu_j$.
If the latter are Majorana particles, 
the $\betabeta$-decay 
experiments can provide constraints
on the absolute scale of
neutrino masses and on the type of neutrino 
mass spectrum. They can also provide unique 
information on the Majorana CP-violation phases 
present in the PMNS neutrino mixing
matrix. The measurement of 
$\meff$ (and $\Sigma$) with 
sufficiently small error 
and sufficiently precise knowledge 
of the values of the 
relevant $\betabeta$-decay nuclear 
matrix elements ($F < 2$)
%
is crucial for obtaining 
significant quantitative information 
on the neutrino mass  and mixing parameters  
from a measurement of 
$\betabeta$-decay half-life.
The remarkably challenging 
physics goal of getting 
evidence for Majorana CP-violation 
in the lepton sector could possibly
be achieved only if {\it in addition}
$\sin^2\theta_{\odot}$ and the
Majorana CP-violation phase
$\alpha_{21}$ have favorable values.

\subsection*{Acknowledgments}

  We would like to thank S.\ M.\ Bilenky for useful discussions.  S.P.
would like to thank the Elementary Particle Physics Sector at SISSA for
kind hospitality during the first stage of this study.
This work was supported in part by the Italian MIUR 
and INFN under the programs on ``Fisica Astroparticellare'' (S.T.P.). 
The work of T.S.\ is supported by a ``Marie
Curie Intra-European Fellowship 
within the 6th European Community
Framework Programme.''


\end{document}